\documentclass[journal]{IEEEtran}
\usepackage{amsmath,amssymb,amsfonts}
\usepackage{algorithm}
\usepackage{algorithmic}
\usepackage{array}
\usepackage[caption=false,font=normalsize,labelfont=sf,textfont=sf]{subfig}
\usepackage[group-separator={,},group-minimum-digits=3]{siunitx}
\usepackage{textcomp}
\usepackage{stfloats}
\usepackage{url}
\usepackage{verbatim}
\usepackage{graphicx}
\usepackage{balance}
\usepackage{mathrsfs}
\usepackage{cite}
\usepackage{xcolor}
\usepackage{booktabs}
\usepackage{threeparttable}

\usepackage[linewidth=1pt]{mdframed}
\usepackage[colorlinks=true,
            linkcolor=blue,
            anchorcolor=blue,
            citecolor=blue]{hyperref}
\begin{document}

\title{  \fontsize{19.5pt}{\baselineskip}\selectfont   
{Simultaneously Exposing and Jamming Covert Communications via Disco Reconfigurable Intelligent Surfaces}
} 
\author{ 
{       Huan~Huang,~\textit{Member,~IEEE}, 
        Hongliang~Zhang,~\textit{Member,~IEEE},
        Yi~Cai,~\textit{Senior~Member,~IEEE}, 
        Dusit~Niyato,~\textit{Fellow,~IEEE},
        A.~Lee~Swindlehurst,~\textit{Fellow,~IEEE},
        and~Zhu~Han~\textit{Fellow,~IEEE}
}
\thanks{  

H.~Huang and Y.~Cai are with the School of Electronic and Information Engineering, Soochow University, Suzhou, Jiangsu 215006, China 
(e-mail: hhuang1799@gmail.com, yicai@ieee.org). 

H.~Zhang is with the State Key Laboratory of Advanced Optical Communication Systems and Networks, 
School of Electronics, Peking University, Beijing 100871, China (email: hongliang.zhang92@gmail.com). 




D.~Niyato is with the College of Computing and Data Science, Nanyang Technological University, Singapore 639798 (e-mail: dniyato@ntu.edu.sg).

A.~L.~Swindlehurst is with the Center for Pervasive Communications and Computing, University of California, Irvine, CA 92697, USA (e-mail: swindle@uci.edu).

Z.~Han is with the Department of Electrical and Computer Engineering at the University of Houston, Houston, TX 77004 USA.  
(email: hanzhu22@gmail.com). 
}
}
\maketitle

\begin{abstract}
Covert communications provide a stronger privacy protection than cryptography and physical-layer security (PLS).
However, previous works on covert communications have implicitly assumed the validity of channel reciprocity, i.e.,
wireless channels remain constant or approximately constant during their coherence time.
In this work, we investigate covert communications in the presence of a disco RIS (DRIS) deployed by the warden Willie,
where the DRIS with random and time-varying reflective coefficients acts as a ``disco ball'',
introducing time-varying fully-passive jamming (FPJ).
Consequently, the channel reciprocity assumption no longer holds.
The DRIS not only jams the covert transmissions between Alice and Bob, 
but also decreases the error probabilities of Willie's detections, 
without either Bob's channel knowledge or additional jamming power.
To quantify the impact of the DRIS on covert communications, 
we first design a detection rule for the warden Willie in the presence of time-varying FPJ introduced by the DRIS.
Then, we define the detection error probabilities, i.e.,
the false alarm rate (FAR) and the missed detection rate
(MDR), as the monitoring performance metrics for Willie's detections, 
and the signal-to-jamming-plus-noise
ratio (SJNR) as a communication performance metric for
the covert transmissions between Alice and Bob.
Based on the detection rule, 
we derive the detection threshold for the warden Willie to detect whether communications between Alice and Bob is ongoing, 
considering the time-varying DRIS-based FPJ.
Moreover, we conduct theoretical analyses of the  FAR and the MDR at the warden Willie, 
as well as SJNR at Bob, and then present unique properties of the DRIS-based FPJ in covert communications. 
We present numerical results to validate the derived theoretical analyses 
and evaluate the impact of DRIS on covert communications.
\end{abstract}

    
\section{Introduction}\label{Intro}
Due to the broadcast nature and superposition properties of wireless channels, 
wireless systems are inherently vulnerable to various malicious attacks~\cite{PLSsur1,DoSsur1,AntiJammingSurv}. 
This issue is particularly critical in the Internet-of-Things (IoT) era, 
where the data transmitted often contains sensitive personal information, 
such as health and location data, 
and in scenarios involving government and military operations, 
where maintaining stealth is essential. 
As a result, research into transmission security and privacy has been advancing rapidly.
Covert communications~\cite{CCRef2}, also known as low probability of detection communications, 
aim to conceal the transmission's existence from adversarial wardens 
by hiding the transmission within environmental noise~\cite{CCRef1,CCRef11}.

Covert communications offer a higher level of privacy protection compared to cryptography and physical-layer security (PLS)~\cite{CCPLS,CCPLS1}
because the warden Willie will not attempt to decode the information contained in the signals 
if he is unaware of the transmission.
In~\cite{CCRef1}, the authors established a fundamental result in covert communications: $o(\sqrt{n})$ bits can be sent in $n$ additive white Gaussian noise (AWGN) channel uses 
while achieving an arbitrarily low probability of detection (LPD)
without knowledge of the noise power on the channel between the transmitter Alice and the warden Willie,
where $o(\sqrt{n})$ represents a non-asymptotically tight upper bound on $\sqrt{n}$.
Furthermore, if a lower bound on the noise power is known, 
up to $\mathcal{O} (\sqrt{n})$ bits can then be sent,
where $\mathcal{O} (\sqrt{n})$ is an asymptotically tight upper bound on $\sqrt{n}$.

Following the work in~\cite{CCRef1}, 
previous studies have introduced other techniques, 
such as non-orthogonal multiple access (NOMA) or Turbo encoding,
to further enhance covert communications
 while guaranteeing transmission performance~\cite{CCNOMA,CCNOMA1,CCTubrbo}.
The works in~\cite{CCRef3,CCRef4,CCRef34,CCRef35} investigated the use of a jammer or adding artificial noise
to further increase the power variation,
thereby enhancing the difficulty of accurate decision-making by the warden Willie.
In addition, \cite{CCRelay1,CCRelay11} demonstrated that the use of a relay can further increase power variations, 
thereby disrupting the warden Willie's detection~\cite{CCRelay2}.
The authors of~\cite{CCSmallScale} proposed exploiting 
the variations in received power due to small scale fading in order to implement covert communications. 

As summarized in Table~\ref{tab2}, the signals transmitted by Alice in covert communications can be effectively concealed from the warden Willie by 
uncertainty in the wireless communication channel, either due to inherent properties such as noise and fading, or factors introduced by Bob and Alice such as relays, coding, or jamming.
    
\begin{table}[ht]
    \footnotesize
    \centering
    \caption{Comparison of Covert Communication Implementations}
    \label{tab2}
    \begin{threeparttable}
    \begin{tabular}{ |c|c|c| }
    \hline
    Implementation               &Requirement                       &Reference\\
    \hline
    Environmental noise          &None                              &\cite{CCRef1,CCRef11} \\
    \hline
    Path loss                    &None                              &\cite{CCSmallScale} \\
    \hline
    Jamming/Artificial noise     &Extra jamming power               &\cite{CCRef3,CCRef4,CCRef34,CCRef35}  \\
    \hline
    Relaying                     &Extra signal processes            &\cite{CCRelay1,CCRelay11,CCRelay2}\\
    \hline
    \end{tabular}
    \end{threeparttable}
\end{table}

Recently, reconfigurable intelligent surfaces (RISs) have been considered as a critical technology to improve wireless communication performance~\cite{PGFun,CHuang,AORIS,IRSsur11,IRSsur3,IRSHLZhang}.
These surfaces consist of numerous elements with reflective coefficients that can be adjusted using simple programmable PIN or varactor diodes~\cite{IRSCuiTJ}.
The integration of RISs into wireless networks significantly enhances their performance 
without substantially increasing power consumption or cost~\cite{EE1,HuangDLRIS}.
The use of RISs in covert communications has already been explored in~\cite{CCRIS1,CCRISNZ1,CCRISNZ2,CCRIS2,CCRISNZ}.
These studies have focused on exploiting one or more RISs for covert communications in systems employing NOMA~\cite{CCRIS1}, 
artificial noise~\cite{CCRISNZ1}, 
finite blocklength coding with variable prior probabilities~\cite{CCRISNZ2}, 
or unmanned aerial vehicles~\cite{CCRIS2,CCRISNZ}.

All existing works on covert communications, with or without RISs,
assume that channel reciprocity in time-division duplex (TDD) wireless channels either holds or is approximately valid.
While such an assumption is normally reasonable, channel reciprocity can be broken in the presence of time-varying ``Disco'' RIS (DRIS)~\cite{MyWCMag}
or RIS employing non-reciprocal connections between their elements~\cite{LeeDRIS,LeeAdd}.
The concept using DRISs to launch 
 fully-passive jamming (FPJ) attacks without relying on either channel knowledge of 
legitimate users or additional jamming power was further developed in~\cite{MyTVT}.
The DRIS coefficients are time-varying and random, acting like a ``disco ball''. 
Consequently, active channel aging (ACA) is introduced, 
invalidating the channel reciprocity of TDD channels even within the channel coherence time~\cite{DIRSTWC,TWCAnti}.
Some works have also exploited DIRSs to disrupt key consistency 
in channel reciprocity-based key generation~\cite{OtherDIRS1}.
 
In this work, we investigate the novel concept of using DRISs to disrupt channel reciprocity in covert communications.
The main contributions are summarized as follows:
\begin{itemize}
\item 
We first present the model of covert communications in the presence of a DRIS, 
whose random and time-varying reflective coefficients introduce FPJ.
The DRIS-based FPJ not only impacts the covert communication between Alice and Bob
but also decreases the detection error probabilities of the warden Willie, 
with neither Alice-Bob channel knowledge nor additional jamming power.
To characterize the impact of the DRIS-based FPJ
we design a detection rule for the warden Willie.
We use the resulting 
 false alarm rate (FAR) and missed detection rate (MDR) as monitoring performance metrics for Willie,
and define the signal-to-jamming-plus-noise ratio (SJNR) as a communication performance metric for covert communications between Alice and Bob.
\item  
To quantify the impact of the time-varying DRIS-based FPJ,
the statistics of the DRIS-influenced channels are first derived.
Based on the derived statistics and the designed detection rule, 
we then determine the detection threshold for the warden Willie to 
decide whether Alice and Bob are transmitting, 
considering the impact of the time-varying DRIS.
Given the detection threshold, closed-form expressions for the FAR and MDR at Willie are derived. 
Furthermore, an asymptotic analysis of the SJNR is conducted to demonstrate the impact of the DRIS-based FPJ on the communications
between Alice and Bob. 
Simulation results are provided to validate the accuracy of the theoretical analyses.
\item  
Based on the detailed theoretical analysis, we present
unique properties of DRIS-based FPJ in covert communications. 
For example, the DRIS not only reduces the detection error probabilities at the warden Willie but also significantly disrupts the transmission between Alice and Bob, even when Willie experiences a missed  detection.
Increasing the transmit power at Alice does not significantly improve communication performance due to the DRIS. 
Instead, it exacerbates the impact of the DRIS-based FPJ on communications between Alice and Bob, and increases Alice's risk of detection by the warden Willie.
Moreover, a DRIS with only 1-bit quantized reflection coefficients is sufficient to enhance the detection
accuracy at the warden and degrade the communication performance between Alice and Bob.
\end{itemize}

The rest of this paper is organized as follows. 
In Section~\ref{Princ}, we model covert communications in the presence of a DRIS, 
and the resulting impact on the wireless channels.
Then, we define the FAR and MDR as performance metrics for  Willie, 
and the SJNR as the performance metric for
the covert transmissions between Alice and Bob.
In Section~\ref{Analysis}, we derive the statistics of the DRIS-based 
channels, and then we determine the detection threshold for the warden Willie to decide whether Alice and Bob are transmitting, 
taking into account the impact of the time-varying DRIS-based FPJ.
Given the detection threshold, closed-form expressions for the FAR and MDR at the warden are derived.  
Moreover, an asymptotic analysis of the SJNR at Bob is conducted.
In Section~\ref{ResDis}, the simulation and theoretical results are compared to validate the derived theoretical analyses
and evaluate the impact of the DRIS.
Finally, conclusions are given in Section~\ref{ResDis}.

\emph{Notation:} 
We employ lowercase bold letters for a vector, e.g., ${\boldsymbol g}$, 
and italic letters for a scalar, e.g., $N_{\rm D}$. 
The operators $(\cdot)^{T}$ and $(\cdot)^{H}$ respectively represent the transpose and the Hermitian transpose, 
and the symbol $|\cdot|$  denotes the absolute value.
Statistical expectation is indicated using $\mathbb{E}\!\left[\cdot\right]$.

\section{System Description}\label{Princ}
In Section~\ref{DRISChannel}, we first present the assumed covert communications scenario in the presence of a DRIS.
Then, we model the wireless channels involved in Section~\ref{ChannelModel}.
In Section~\ref{PerformanceMetrics}, we describe the detection rule performed by the warden Willie to 
monitor the covert communications between Alice and Bob, 
and we define the FAR and MDR.
In Section~\ref{SJNRAB}, we define the SJNR as the communication performance metric 
used to
evaluate the impact of the DRIS-based FPJ on the covert transmissions between Alice and Bob.  
\subsection{Covert Communications in the Presence of a DRIS}\label{DRISChannel}
\begin{figure}[!t]
    \centerline{\includegraphics[scale=0.078]{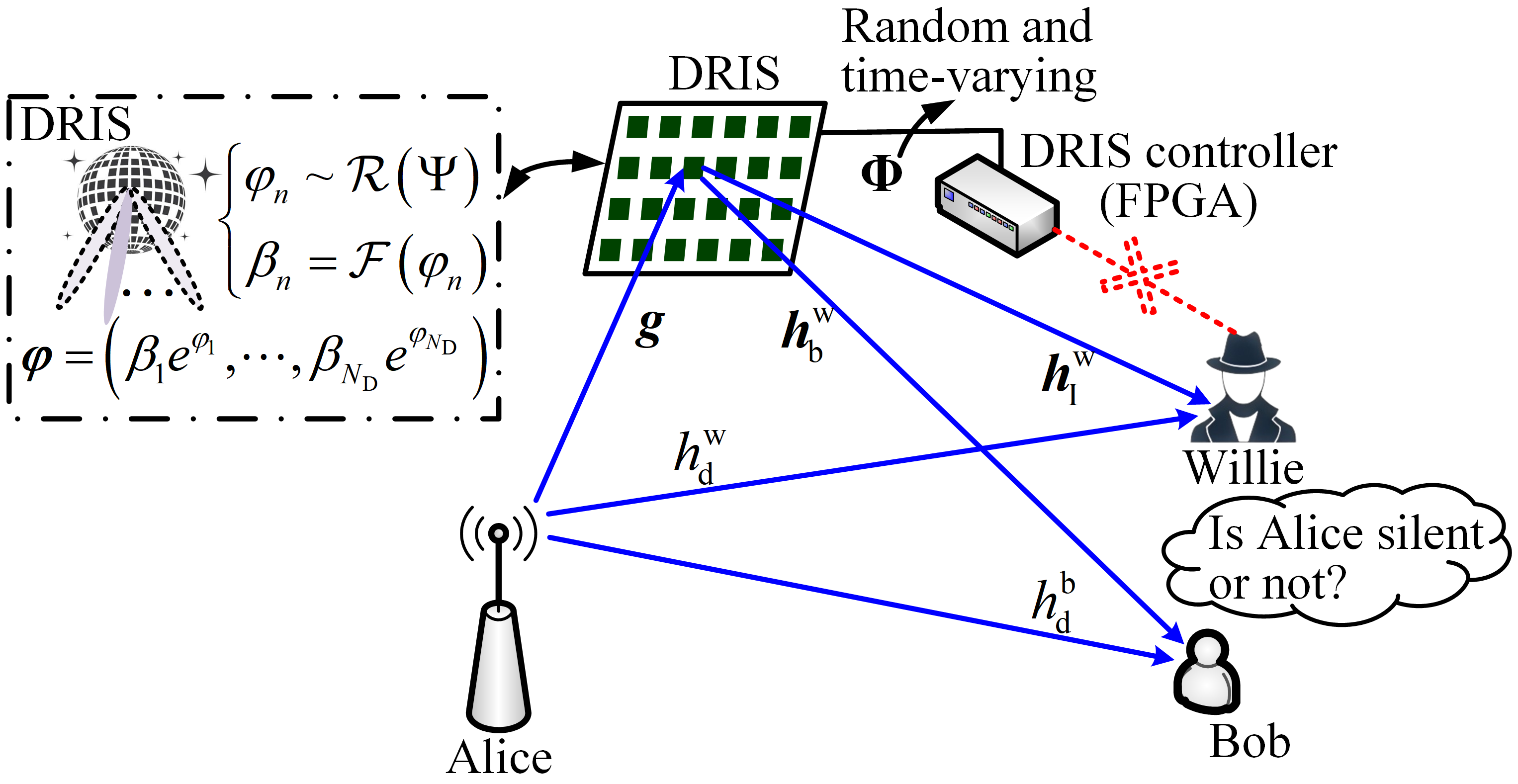}}
    \caption{Covert communications in the presence of a disco reconfigurable intelligent surface (DRIS),
    {\textcolor[rgb]{0.00,0.00,1.00}{whose time-varying and random reflection coefficients are
     generated by a DRIS controller 
    without any connection or coordination with Willie}}.}
    \label{fig1}
\end{figure}
Fig.~\ref{fig1} schematically illustrates a covert communication system in the presence of a DRIS.
In covert communications, Alice aims to covertly transmit messages to Bob  avoiding detection by the warden Willie.
In this work, we investigate the problem of determining the probability of Willie to correctly detect whether Alice and Bob are silent or not, 
without any coordination with them.
Therefore, we assume that Willie has no channel state information (CSI) for Bob, including his location.

To more accurately detect whether Alice is transmitting, 
the warden employs an $N_{\rm D}$-element DRIS ($N_{\rm D} = N_{{\rm D},h} \times N_{{\rm D},v}$) with coefficients tuned by a simple programmable PIN~\cite{IRSCuiTJ}.
{\textcolor[rgb]{0.00,0.00,1.00}{We use a DRIS controller implemented 
using  field-programmable gate arrays (FPGAs)
to control the simple programmable PIN. 
The DRIS controller operates autonomously and does not require any connection or coordination with Willie.}}
The ON/OFF behavior of the PIN  only allows for the implementation of discrete DRIS coefficients obtained by
$b$-bit quantization of the reflection coefficient phase shifts and amplitudes.
More specifically, the possible phase shifts and amplitudes of the DRIS are denoted as
$\Psi  = \{ {{\phi _1},{\phi _2}, \cdots, {\phi _{{2^b}}}} \}$ and  $\Omega = \{ {{\alpha_1}, {\alpha_2}, \cdots, {\alpha_{{2^b}}}} \}$,
respectively.     

Similar to previous work in~\cite{MyWCMag, MyTVT, DIRSTWC, TWCAnti}, 
the reflection coefficients $\varphi_n(t)$ and $\beta_n(t)$ of the $n$-th DRIS element ($n=1,\cdots, N_{\rm D}$) are randomly generated and time-varying, 
and are assumed to be independent and identically distributed (i.i.d.) for different $n$.
{\textcolor[rgb]{0.00,0.00,1.00}{According to~\cite{TWCAnti}, the DRIS coefficients only need to be changed a few times within each
channel coherence interval to age the CSI.}}
Consequently, the passive beamforming of the DRIS can be expressed as ${\bf \Phi}(t) = {\rm{diag}}\left({\boldsymbol \varphi }(t)\right)$
where the DRIS reflection vector is denoted as ${\boldsymbol \varphi }(t) = \left[\beta_1(t)e^{\varphi_1(t)}, \cdots,\beta_{N_{\rm D}}(t)e^{\varphi_{N_{\rm D}}(t)} \right]$.
In practice, $\beta_n(t)$ is a function of $\varphi_n(t)$, i.e.,  $\beta_n(t)= \mathcal{F}\!\left(\varphi_n(t)\right) \in \Omega$~\cite{PGFun}.

Based on the covert communication model in~\cite{CCRIS1,CCRef2}, 
the $m$-th sample at the warden Willie over a given channel coherence interval can be expressed as
\begin{alignat}{1}
    \nonumber 
    &{y_{\rm{w}}(m)} = \\
     &\;\; \left\{\!\!\! \begin{array}{l}
         {\underbrace {{\frac{{h^{\rm{w}}_{{\rm{d}}} }s(m) }{{{\mathcal{L}^{\frac{{{\nu^{\rm{w}}_{{\rm{d}}}}}}{2}}}}}}}_{\rm{Direct\;link}}} \!+\!\!\!
         {\underbrace {{\frac{{h^{\rm{w}}_{{\rm{D}}}(m) }s(m)}{{  {\mathcal{L}^{\frac{{{\nu _{{\rm{g}}}}}}{2}}} {\mathcal{L}^{\frac{{{\nu^{\rm{w}}_{{\rm{I}}}}}}{2}}} } }}}_{\rm{DRIS-Based\;link}}} \!\!+
         {n_{{\rm w}}(m)}, \;{\rm{transmitting}}, \\
         {}\\
         {n_{{\rm w}}(m)}, \;\;\;\;\;\;\;\;\;\;\;\;\;\;\;\;\;\;\;\;\;\;\;\;\;\;\;\;\;\;\;\;
         \;\;\;\;\;\;\;\;\;\;\; {\rm{silent}},
         \end{array} \right. 
     \label{CCObSigWS}
 \end{alignat}
where $s(m)$ represents a covert symbol transmitted by Alice,
and $n_{\rm w}(m)$ is AWGN with zero mean and 
 variance $\delta_{\rm{w}}^2$, i.e., $n_{\rm w}(m)\sim {\mathcal{CN}}\!\left(0,\delta_{\rm{w}}^2\right)$.
Similar to~\cite{CCRef1,CCRef3,CCRef4},
we assume that the covert symbols transmitted by Alice  
follow the complex Gaussian distribution with zero mean and variance $P_0$, i.e.,
$s(m)\sim {\mathcal{CN}}\!\left(0,P_0\right)$,
where $P_0$ is the transmit power at Alice. 
 
In~\eqref{CCObSigWS}, ${h^{\rm{w}}_{{\rm{d}}}}$ represents the direct channel between Alice and Willie, 
and ${h^{\rm{w}}_{{\rm{D}}}(m) }$ denotes the cascaded DRIS-based channel between Alice and Willie at the $m$-th sampling time.
The cascaded DRIS-based channel ${\boldsymbol h}^{\rm w}_{\rm D}(m)$ can be expressed as
\begin{equation}
    { h}^{\rm w}_{\rm D}(m) = {{\boldsymbol g }}
     {{\rm{diag}}\!\left({\boldsymbol \varphi }(m) \right)} {{\boldsymbol{h}}^{{\rm w}}_{{\rm I}} },
    \label{hDEx}
\end{equation}
where ${{\boldsymbol g }} \in {\mathbb{C}}^{ 1 \times {N_{\rm D}}}$ denotes the channel between Alice and the DRIS,
 and ${{\boldsymbol{h}}^{{\rm w}}_{{\rm I}}} \in {\mathbb{C}}^{ {N_{\rm D}} \times 1}$ denotes the channel between the DRIS and Willie.
 The factors
 ${{\mathcal{L}^{\frac{{{\nu^{\rm{w}}_{{\rm{d}}}}}}{2}}}}$,
 ${{\mathcal{L}^{\frac{{{\nu_{{\rm{g}} }}}}{2}}}}$,
 and ${{\mathcal{L}^{\frac{{{\nu^{\rm{w}}_{{\rm{I}}}}}}{2}}}}$
 represent
 the large-scale fading coefficients for ${h^{\rm{w}}_{{\rm{d}}}}$, ${\boldsymbol{g}}$,
 and ${{\boldsymbol{h}}^{\rm w}_{{\rm I}}}$
 with corresponding path loss exponents of ${\nu^{\rm{w}}_{{\rm{d}}}}$, ${\nu _{{\rm{g}}}}$, and
 ${\nu^{\rm{w}}_{{\rm{I}}}}$, respectively.
In addition, we assume that Willie takes $M \ge 2$ samples within the channel coherence time,
resulting in the vector
${\boldsymbol{y}}_{\rm w} = \left[{y}_{\rm w}(1),{y}_{\rm w}(2),\cdots,{y}_{\rm w}(M)\right]^T$.

The $m$-th signal sample received by Bob is given by 
\begin{equation}
    {y_{\rm{b}}(m)} = 
    {\underbrace {{\frac{{h^{\rm{b}}_{{\rm{d}}}}s(m)}{{{\mathcal{L}^{\frac{{{\nu^{\rm{b}}_{{\rm{d}}}}}}{2}}}}}}}_{\rm{Direct\;link}}} + 
         {\underbrace {{\frac{{h^{\rm{b}}_{{\rm{D}}}(m)}s(m) }{{  {\mathcal{L}^{\frac{{{\nu_{{\rm{g}}}}}}{2}}} {\mathcal{L}^{\frac{{{\nu^{\rm{b}}_{{\rm{I}}}}}}{2}}} } }}}_{\rm{DRIS-Based\;link}}} + 
         {n_{{\rm b}}(m)}, 
     \label{CCObSigBS}
\end{equation}
where ${h^{\rm{b}}_{{\rm{d}}}}$ denotes the direct channel between Alice and Bob,
${h^{\rm{b}}_{{\rm{D}}}(m)}$ represents the cascaded DRIS-jammed channel between Alice and Bob
at sample $m$,
and $n_{\rm b}(m)$ is AWGN satisfying $n_{\rm b} \sim {\mathcal{CN}}\!\left(0,\delta_{\rm{b}}^2\right)$.
In~\eqref{CCObSigBS}, ${{\mathcal{L}^{\frac{{{\nu^{\rm{b}}_{{\rm{d}}}}}}{2}}}}$ and 
${{\mathcal{L}^{\frac{{{\nu^{\rm{b}}_{{\rm{I}}}}}}{2}}}}$
 represent the large-scale fading coefficients for ${h^{\rm{b}}_{{\rm{d}}}}$ 
and ${{\boldsymbol{h}}^{\rm b}_{{\rm I}}}$, respectively. 
The cascaded DRIS-jammed channel ${ h}^{\rm b}_{\rm D}(m)$ is expressed as
\begin{equation}
    { h}^{\rm b}_{\rm D}(m) = {{\boldsymbol g }}
     {{\rm{diag}}\!\left({\boldsymbol \varphi }(m) \right)} {{\boldsymbol{h}}^{{\rm b}}_{{\rm I}} },
    \label{hDBob}
\end{equation}
where ${{\boldsymbol{h}}^{{\rm b}}_{{\rm I}}} \in {\mathbb{C}}^{ {N_{\rm D}} \times 1}$ denotes the channel between the DRIS and Bob.


\subsection{Channel Model}\label{ChannelModel}
In this work, we consider a scenario where the DRIS is deployed close to Alice$\footnote{\textcolor[rgb]{0.00,0.00,1.00}{Many existing performance-enhancing RIS-aided communication systems assume that
legitimate RISs are deployed close to users in order to maximize system performance~\cite{IRSdeployment}. 
However, in the covert communication scenario presented here, we make
the more robust assumption that the independent DIRS controller does not
have any information about Bob, such as his location. The
location of Alice is fixed, and thus we assume that the DIRS is deployed
near Alice. Our deployment strategy is informed by the conclusion given
in~\cite{IRSdeployment}, which makes the impact of the DIRS as large as possible.} }$, 
while the warden Willie and Bob are positioned farther away, \textcolor[rgb]{0.00,0.00,1.00}
{beyond the near-field range of the DRIS}.
\textcolor[rgb]{0.00,0.00,1.00}{
Deploying the DRIS close to Alice may increase the risk of being detected.
Fortunately, an interesting property of RISs is their passive nature, which means that an RIS
cannot transmit/receive, or process signals~\cite{IRSCuiTJ}.
In addition, an RIS can be easily hidden in the
environment by disguising it, for instance, by embedding it in a glass structure. In practical scenarios,
an RIS can also be mounted on walls or integrated into existing infrastructure. 
Therefore, the DRIS can be set up ahead of time and hidden using the aforementioned camouflage techniques.}

The direct \textcolor[rgb]{0.00,0.00,1.00}{Alice-Bob} and Alice-Willie channels ${{ {h}}^{\rm b}_{{\rm d}}}$ 
and ${{ {h}}^{\rm w}_{{\rm d}}}$,
and the DRIS-Bob and DRIS-Willie channels ${{\boldsymbol{h}}^{\rm b}_{{\rm I}}}$ 
and ${{\boldsymbol{h}}^{\rm w}_{{\rm I}}}$
are modeled based on the far-field assumption, 
with elements that are assumed to be i.i.d. Gaussian random variables defined as~\cite{BookFarFeild}. 
\begin{alignat}{1}
    &{{h}^{\rm b}_{\rm{d}}},{{h}^{\rm w}_{\rm{d}}} \sim  {\mathcal{CN}}\!\left(0,1\right) ,\label{Hdbkeq}\\
    &{{\boldsymbol{h}}^{\rm b}_{{\rm I}}}, {{\boldsymbol{h}}^{\rm w}_{{\rm I}}} 
    \sim  {\mathcal{CN}}\!\left({{\bf 0}_{N_{\rm D}}},{{\bf I}_{N_{\rm D}}}\right),
    \label{Hdkeq}
\end{alignat}
where ${{\bf 0}_{N_{\rm D}}}$ is the ${N_{\rm D}} $-dimensional
zero vector and
${{\bf I}_{N_{\rm D}}}$ is the ${N_{\rm D}} $-dimensional identity matrix.

The DRIS (and RIS in general) typically require a large number of reflective elements to mitigate the significant impact 
of multiplicative large-scale fading~\cite{MyWCMag,DIRSTWC,TWCAnti}. 
Therefore, the Alice-DRIS channel ${{\boldsymbol{g}}}$ is generated based on 
a near-field model~\cite{NearfieldMo1}:
\begin{equation}
        {{\boldsymbol{g}}} =   \sqrt {\frac{{{\varepsilon_{\rm{g}}}} }{{1\!+\!{\varepsilon_{\rm{g}}}}}} {{ {{\boldsymbol{g}}}}^{{\rm{LOS}}}}  \!+\! 
        \sqrt {\frac{1}{{1\!+\!{\varepsilon_{\rm{g}}}}}} {{ {{{\boldsymbol{g}}}}}^{{\rm{NLOS}}}},
\label{Ricianchan}
\end{equation}
where ${{\varepsilon_{\rm{g}}}}$ represents the Rician factor 
for ${{\boldsymbol{g}}}$.  
In~\eqref{Ricianchan},
the elements of the non-line-of-sight (NLoS) component ${{ {{\boldsymbol{g}}}}^{{\rm{NLOS}}}}$ 
follow Rayleigh fading~\cite{BookFarFeild}.
On the other hand,  the elements of the line-of-sight (LoS) component ${{ {\boldsymbol{g}}}^{{\rm{LOS}}}}$ are given by~\cite{NearfieldMo1}
\begin{equation}
    \left[{{ {\boldsymbol{g}}}^{{\rm{LOS}}}}\right]_{r}  
    =  {e^{ \!- j\frac{{2\pi }}{\lambda }\left( {{d_{r}} - {d_0}} \right)}},
    r = 1,\cdots, N_{\rm D},
    \label{GLOS}
\end{equation}
where $\lambda$ is the wavelength of the covert signals, 
${d_{r}}$ and ${d_0}$ represent the distance between Alice's antenna and the $r$-th DIRS element,
 and the distance between this antenna and the centre (origin) of the DIRS, 
respectively.

\subsection{Detection by Willie in the Presence of a DRIS}\label{PerformanceMetrics}
In covert communications, the warden Willie monitors the wireless channels to determine whether Alice and Bob are transmitting.
In particular,  Willie deduces from the samples in~\eqref{CCObSigWS} 
which of the following two events has occurred:
Alice and Bob are communicating (${\mathcal H}_1$), or Alice is silent (${\mathcal H}_0$).
The following two error probabilities are widely used to measure the detection performance:
 ${\mathbb{P}}\left({{\mathcal{H}}_1}|{{\mathcal{H}}_0}\right)$, which is the probability that 
 Willie concludes Alice and Bob are communicating when they are not (the FAR), 
 and ${\mathbb{P}}\left({{\mathcal{H}}_0}|{{\mathcal{H}}_1}\right)$, which is the probability that 
Willie concludes Alice and Bob are silent when they are communicating (the MDR).

Since existing covert communication work relies almost entirely on 
the assumption that TDD channel reciprocity holds, 
the elements of the received sample vector ${\boldsymbol{y}}_{\rm w}$ are i.i.d. 
Consequently, Willie can use the total power of ${\boldsymbol{y}}_{\rm w}$
as a detection statistic to decide whether Alice and Bob are communicating~\cite{CCRIS1,CCRIS2,CCRef4,CCRelay1,CCRelay11,CCRelay2,CCSmallScale}.
However, with the introduction of the DRIS, the basic assumption of channel reciprocity no longer applies. 
Consequently, the elements of the received sample vector ${\boldsymbol{y}}_{\rm w}$
are no longer i.i.d., even though they are received within the same channel coherence interval.
The DRIS introduces the signal through an additional random and time-varying 
channel  $h^{\rm w}_{\rm D} (t)$
to the observation in~\eqref{CCObSigWS}, when Alice is transmitting to Bob.
\textcolor[rgb]{0.00,0.00,1.00}{Unlike~\cite{CCRIS1,CCRIS2,CCRef4,CCRelay1,CCRelay11,CCRelay2,CCSmallScale},
we show below that, for covert communications in the presence of a DRIS, using the total received power 
 of ${\boldsymbol{y}}_{\rm w}$
as a test statistic to decide whether Alice is transmitting is analytically intractable and less practical, 
because the corresponding optimal detection threshold does not admit a closed-form expression.}

To take into account the presence of the DRIS,
we propose that Willie use the following \textcolor[rgb]{0.00,0.00,1.00}{decision region $\mathcal{S}$}:
\begin{alignat}{1}
    \nonumber
    & {\mathcal{S}} \\
    &= \!\left\{\!\! {\left. {\boldsymbol{y}_{\rm w}} \right|\mathop  \cup \limits_{ \!{i_1} \!<  \cdots < {i_N}  \!} \!\!\left(\! {{{\left| {{y_{\rm{w}}}\left( {{i_1}} \right)} \right|}^2} \!\ge \! \varepsilon (i_1) \!\cap  \! \cdots \!\cap\! {{\left| {{y_{\rm{w}}}({{i_N}})} \right|}^2} \!\ge\! \varepsilon( {{i_{\!N}}})} \! \right)} \!\!\right\}, 
    \label{decisionrule1}
\end{alignat}
where $1 \le {i_1} <  \cdots < {i_N} \le M$, 
and ${\varepsilon}(m)$ denotes the \textcolor[rgb]{0.00,0.00,1.00}{optimal} detection threshold for the $m$-th 
component,
\textcolor[rgb]{0.00,0.00,1.00}{which can be determined using the likelihood ratio test (LRT).
We will see below that
 ${\left| {{y_{\rm{w}}}\left( {{m}} \right)} \right|}^2$ is a sufficient statistic.
}

Based on~\eqref{decisionrule1}, the detection made by the warden Willie can be expressed mathematically as
\begin{alignat}{1}
    &{{\mathcal{H}}_1}: \; {\boldsymbol{y}_{\rm w}} \in  {\mathcal{S}}, \label{decisionrulexx}\\
    &{{\mathcal{H}}_0}: \; {\boldsymbol{y}_{\rm w}} \notin {\mathcal{S}}. 
    \label{decisionrule}
\end{alignat}
Note that the time-varying detection thresholds ${\varepsilon}(m), m = 1,\cdots, M$ should be designed
by considering the influence of the DRIS
to ensure detection accuracy. 
The FAR and MDR are correspondingly expressed as
\begin{alignat}{1}
    p_{\rm{F}} & = {\mathbb{P}}\! \left( {{\mathcal{H}}_1}|{{\mathcal{H}}_0} \right) = {\mathbb{P}}\! \left( {\boldsymbol{y}}_{\rm{w}} \in {\mathcal{S}}|{{\mathcal{H}}_0} \right) , \label{FARProb}\\
    p_{\rm{M}} & = {\mathbb{P}}\! \left( {{\mathcal{H}}_0}|{{\mathcal{H}}_1} \right) = {\mathbb{P}}\! \left( {\boldsymbol{y}}_{\rm{w}} \notin  {\mathcal{S}}|{{\mathcal{H}}_1} \right)  \label{MDRProb}.
\end{alignat}

\subsection{Signal-to-Interference-Plus-Noise Ratio at Bob}\label{SJNRAB}
Based on~\eqref{CCObSigBS}, the received signals at Bob are subject to jamming by the DRIS 
due to  $h^{\rm b}_{\rm D}(t)$.
To characterize the impact of the DRIS on Bob$\footnote{\textcolor[rgb]{0.00,0.00,1.00}{
The approach presented in the paper can easily be extended to the case of multiple DRISs. 
In such a scenario, each DRIS chooses its own random time-varying reflection coefficients, 
further increasing the ACA effect on communications between Alice and Bob.}}$, 
we use the SJNR defined by~\cite{TWCAnti}
\begin{equation}
    {\eta _{\rm b}} =  
    \frac{\frac{ \mathbb{E} \!\left[\left|{h^{\rm{b}}_{{\rm{d}}}}s(m)\right|^2\right] }{{{\mathcal{L}^{{{{\nu^{\rm{b}}_{{\rm{d}}}}}}}}}}}
    { \frac{ \mathbb{E} \!\left[\left|{h^{\rm{b}}_{{\rm{D}}}(m)}s(m)\right|^2\right] }{{{\mathcal{L}^{{{{\nu _{{\rm{g}}}}}}}}}{{\mathcal{L}^{{{{\nu^{\rm{b}}_{{\rm{I}}}}}}}}}} + \delta _{\rm{b}}^2}.
    \label{eqSLNR}
\end{equation}
Consequently, the achievable rate at Bob is given by
$R_{\rm b} = {\log _2}\left( {1 + {\eta _{\rm b}} } \right)$.
 
From~\eqref{eqSLNR}, one can see that the covert transmissions between Alice and Bob are jammed by 
the time-varying DRIS,
even though no jamming power is introduced, and Willie does not require CSI for Bob.
In particular, due to the time-varying DRIS reflection coefficients ${\boldsymbol \varphi }(t)$,
the cascaded DRIS-jammed channel ${h^{\rm{b}}_{{\rm{D}}}}(t)=
{{\boldsymbol g }}
     {{\rm{diag}}\!\left({\boldsymbol \varphi }(t) \right)} {{\boldsymbol{h}}^{{\rm b}}_{{\rm I}} }$ is 
time-varying even within the channel coherence time.
The vector ${\boldsymbol \varphi }(t)$ is randomly generated by Willie and is unknown to Alice and Bob.
As a result, ACA interference$\footnote{The introduced ACA interference is different from interference due to channel aging (CA)~\cite{ChanAge},
which arises from time variations in the RF propagation and computational delays between the time the channels are learned at the legitimate transmitter and when they are used for precoding.}$ is introduced by breaking the channel reciprocity, 
which decreases the SJNR at Bob.
This implies that the deployment of the DRIS by the warden Willie not only 
improves his detection performance but also degrades the communication performance between Alice and Bob.

\section{Asymptotic Analysis of Covert Communication Performance}\label{Analysis}
In Section~\ref{OBiDet}, the statistics of the DRIS-based channels are first derived 
to quantify the impact of the time-varying DRIS-based FPJ on covert communications. 
Then, we determine the detection threshold for the warden Willie to decide whether Alice and Bob are
transmitting, considering the impact of the time-varying DRIS. 
In Section~\ref{BERDec}, given the detection threshold,
closed-form expressions for the detection error probabilities at Willie are derived.
In Section~\ref{SNRDec}, an asymptotic analysis of the SJNR is
conducted to illustrate the impact of the DRIS on the covert  transmissions between Alice and Bob.

\subsection{Detection Error Probability in the Presence of a DRIS}\label{OBiDet}
In order to determine the detection thresholds in~\eqref{decisionrule1}, 
we first derive the statistics of the cascaded
DRIS-based channel ${{\boldsymbol{h}}^{\rm{w}}_{{\rm{D}}} (t)}$ between Alice and Willie.

\newtheorem{proposition}{Proposition}
\begin{proposition}
    \label{Proposition1}
    The random and time-varying DRIS-based channel ${{h}^{\rm{w}}_{{\rm{D}}}(t)}$
    converges in distribution to a complex Gaussian random variable as $N_{\rm D} \to \infty$, i.e.,
    \begin{equation}
        \frac{{h}^{\rm{w}}_{{\rm{D}}} (t)}{{{\mathcal{L}^{{{\frac{\nu_{{\rm{g}}}}{2}}}}}}{{\mathcal{L}^{{{\frac{\nu^{\rm{w}}_{{\rm{d}}}}{2}}}}}}}
        \mathop  \to \limits^{\rm{d}} \mathcal{CN}\!\!\left( {0,  {  \frac{{N\!_{\rm D}}{\overline \alpha}}{{{\mathcal{L}^{{{{\nu_{{\rm{g}}}}}}}}}{{\mathcal{L}^{{{{\nu^{\rm{w}}_{{\rm{d}}}}}}}}}} } } \right), 
        \label{HDSta}
    \end{equation}
where $\overline \alpha =  \mathbb{E} \!\left[\left|\beta_{n}(t) \right|^2\right]= \sum\nolimits_{i = 1}^{{2^b}} {{{\rm P}_i}\alpha _i^2}$, 
and  ${{\rm P}_i}$ is the probability of 
the phase shift ${\varphi _r}(t)$ taking the $i$-th value of $\Phi$,
 i.e., ${{\rm P}_i} = \mathbb{P}\!\left({\varphi _r}(t)= {\phi _i}\right)$.
 For ease of analysis, we assume that the discrete phases chosen at each DRIS element
 are i.i.d. according to a uniform distribution.
{\textcolor[rgb]{0.00,0.00,1.00}{Note that if
all DRIS phase shifts are independently and identically selected from a discrete
set, the conclusion in Proposition~\ref{Proposition1} still holds.
If all DRIS phase shifts are independently but not identically selected from
a discrete set, the conclusion in Proposition~\ref{Proposition1} holds only if the Lyapunov condition is
satisfied.}}
\end{proposition}

\begin{IEEEproof}
    See Appendix~\ref{AppendixA}.
\end{IEEEproof}

To give the optimal detection threshold $\varepsilon$ for Willie,
the distribution of the Willie's received signal is derived.
The $m$-th sample ${ {y}}_{\rm w}(m)$ received at  Willie under hypothesis $\mathcal{H}_1$
 is given by
\begin{equation}
    {{ y}_{\rm{w}}} (m) =  
    {{\frac{{h^{\rm{w}}_{{\rm{d}}} }s(m) }{{{\mathcal{L}^{\frac{{{\nu^{\rm{w}}_{{\rm{d}}}}}}{2}}}}}}} + 
    {{\frac{{h^{\rm{w}}_{{\rm{D}}} (m)}s(m)}{{  {\mathcal{L}^{\frac{{{\nu _{{\rm{g}}}}}}{2}}} {\mathcal{L}^{\frac{{{\nu^{\rm{w}}_{{\rm{I}}}}}}{2}}} } }}} +
    {n_{{\rm w}}(m)}.
     \label{CCObSigWS1}
 \end{equation}
The direct channel ${ h}^{\rm w}_{\rm d}$ is 
constant over the channel coherence interval. Thus, 
based on Proposition~\ref{Proposition1},  
we have the following result:
\begin{alignat}{1}
    \nonumber
    c_{\rm w}(m) &= {{\frac{{h^{\rm{w}}_{{\rm{d}}} } }{{{\mathcal{L}^{\frac{{{\nu^{\rm{w}}_{{\rm{d}}}}}}{2}}}}}}}
    + {{\frac{{h^{\rm{w}}_{{\rm{D}}} (m)} }{{  {\mathcal{L}^{\frac{{{\nu _{{\rm{g}}}}}}{2}}} {\mathcal{L}^{\frac{{{\nu^{\rm{w}}_{{\rm{I}}}}}}{2}}} } }}} \\
    & \mathop  \to \limits^{\rm{d}} \mathcal{CN}\!\left( {{{\frac{{h^{\rm{w}}_{{\rm{d}}} } }{{{\mathcal{L}^{\frac{{{\nu^{\rm{w}}_{{\rm{d}}}}}}{2}}}}}}},
    {  \frac{{N\!_{\rm D}}{\overline \alpha}}{{{\mathcal{L}^{{{{\nu_{{\rm{g}}}}}}}}}{{\mathcal{L}^{{{{\nu^{\rm{w}}_{{\rm{d}}}}}}}}}} } }\right).
    \label{CwGauss}
\end{alignat}


However, in~\eqref{CCObSigWS1}, 
the covert message $s(m)$ also follows a complex Gaussian distribution, which is independent of $c_{\rm w}(m)$.
The product of two independent complex Gaussian variables
is not Gaussian~\cite{CNN}.
For two independent random variables $X$ and $Y$, 
with probability density functions (PDFs)
$f_X(x)$ and $f_Y(y)$, respectively, the PDF of $Z=XY$  can be computed using:
\begin{equation}
{f_Z}\!\left( z \right) = \int_{ \mathbb{C} }  {{f_X}\left( x \right)} {f_Y}\!\left( {\frac{z}{x}} \right)\frac{1}{{\left| x \right|^2}}d^2x.
    \label{XYPDF}
\end{equation}
According to~\eqref{XYPDF}, the PDF of 
$e_{\rm w}(m) = c_{\rm w}(m)s(m)$ is given by
\begin{alignat}{1}
    \nonumber
    &{f_{E_{\rm w}}} ( {\left. {{e_{\rm{w}}(m)}} \right|{\mathcal{H}_1}}  ) \\
    \nonumber
    &= \!\int_{ - \infty }^{ + \infty }\!\! \frac{1}{{ \!{  \frac{{\pi N\!_{\rm D}}{\overline \alpha}}{{{\mathcal{L}^{{{{\nu_{{\rm{g}}}}}}}}}{{\mathcal{L}^{{{{\nu^{\rm{w}}_{{\rm{I}}}}}}}}}} } \! }}
    {e^{ \!\! -\! \frac{{\left| \!{{c_{\rm{w}}}\left( m \right)} - {{\frac{{h^{\rm{w}}_{{\rm{d}}} } }{{{\mathcal{L}^{\frac{{{\nu^{\rm{w}}_{{\rm{d}}}}}}{2}}}}}}}\!\right|^2}}
    {  \frac{{N\!_{\rm D}}{\overline \alpha}}{{{\mathcal{L}^{{{{\nu_{{\rm{g}}}}}}}}}{{\mathcal{L}^{{{{\nu^{\rm{w}}_{{\rm{I}}}}}}}}}} } }}
    \!\! \frac{1}{{\pi \!{P_0}}}{e^{ - \frac{{{{\left|\! {\frac{{{e_{\rm{w}}}(m)}}{{{c_{\rm{w}}}(m)}}} \!\right|}^2}}}{{{P_0}}}}}{\left| \!{\frac{1}{{{c_{\rm{w}}}\!(m)}}}\! \right|^2}d{c_{\rm{w}}}\!(m) \\ 
    & = \frac{4\!\left| e_{\rm w}\!(m)\right|}{  \frac{{P_0 N\!_{\rm D}}{\overline \alpha}}{{{\mathcal{L}^{{{{\nu_{{\rm{g}}}}}}}}}{{\mathcal{L}^{{{{\nu^{\rm{w}}_{{\rm{I}}}}}}}}}} }
    e^{\!-\kappa^2_{\rm e}}\sum\limits_{n = 0}^{ + \infty } {{\!\!{\left( \!{\frac{1}{{n!}}} \!\right)}^{\!2}}}
\!\!\left(\!\frac{\kappa^2_{\rm e}\left| e_{\rm w}\!(m)\right|}{  \frac{{P_0 N\!_{\rm D}}{\overline \alpha}}{{{\mathcal{L}^{{{{\nu_{{\rm{g}}}}}}}}}{{\mathcal{L}^{{{{\nu^{\rm{w}}_{{\rm{I}}}}}}}}}} }\!\right)^{\! n}
\!K_{n}\!\!\!\left(\!\frac{2\left| e_{\rm w}\!(m)\right|}{\sqrt{  \frac{{P_0 N\!_{\rm D}}{\overline \alpha}}{{{\mathcal{L}^{{{{\nu_{{\rm{g}}}}}}}}}{{\mathcal{L}^{{{{\nu^{\rm{w}}_{{\rm{I}}}}}}}}}} }}\!\right),
    \label{ewPDF}
\end{alignat}
where $\kappa_{\rm e}$ is defined as $\frac{\left|h^{\rm w}_{\rm d}\right|^2{{{\mathcal{L}^{{{{\nu_{{\rm{g}}}}}}}}}{{\mathcal{L}^{{{{\nu^{\rm{w}}_{{\rm{I}}}}}}}}}}}{{N\!_{\rm D}}{\overline \alpha}{{\mathcal{L}^{{{{\nu^{\rm{w}}_{{\rm{d}}}}}}}}}}$~\cite{CNN},
and $K_n(\cdot)$ represents the $n$-order modified Bessel function of the second kind.

Let $Z$ and $V$ be two independent random variables with PDFs $f_Z(z)$ and $f_V(v)$, respectively. 
The PDF of $W = Z + V$
is given by
\begin{equation}
    {f_W}\left( w \right) = \int_{ \mathbb{C} } \! {{f_Z}\left( z \right)} {f_V}\left( {w - z} \right)d^2z.
    \label{WVPDF}
\end{equation}
In~\eqref{CCObSigWS1}, the PDF of the AWGN $n_{\rm w}(m)$ can be expressed as
\begin{equation}
    f_{N_{\rm w}}\!\left({  { n_{\rm w}(m)} } \right)   
    =\frac{1}{{\pi \delta _{\rm{w}}^2}}{e^{ - \frac{{\left| {{y_{\rm{w}}} ( m )} \right|^2}}{{\delta _{\rm{w}}^2}}}}.
    \label{nwPDFH1}
\end{equation}
Substituting~\eqref{ewPDF} and~\eqref{nwPDFH1} to~\eqref{WVPDF},
we have 
\begin{alignat}{1}
    \nonumber
    &f_{Y_{\rm w}}\!\left( {\left. { y_{\rm w}(m)} \right|{\mathcal{H}_1}}  \right) \\
    &= \!\!\int_{\mathbb{C} } \!{{f_{{N_{\rm{w}}}}}\!\left( {{y_{\rm w}(m)} - {e_{\rm{w}}\!(m)}} \right)} {f_{{E_{\rm{w}}}}}\!\left( {\left. {{e_{\rm{w}}(m)}} \right|{\mathcal{H}_1}} \right){d {{e_{\rm{w}}}\!(m)}} \label{yPDF1}\\
      &=\!\!\int_{\mathbb{C} } \!{{\frac{1}{{\pi \delta _{\rm{c}}^2}}{e^{ - \frac{{{{\left| {{y_{\rm w}(m)} - {e_{\rm{w}}\!(m)}} \right|}^2}}}{{\delta _{\rm{c}}^2}}}}} 
      \frac{{2\!\left| {e_{\rm w}}\!(m) \right|}}{{\pi \frac{{{P_0}{N_{\rm{D}}}\overline \alpha  }}{{{\mathcal{L}^{{\nu_{\rm{g}}}}}{\mathcal{L}^{\nu_{\rm{I}}^{\rm{w}}}}}}}}{K_0}\!\!\left( {\frac{{\left| {e_{\rm w}\!(m)} \right|}}{{\sqrt {\frac{{{P_0}{N_{\rm{D}}}\overline \alpha  }}{{{\mathcal{L}^{{\nu_{\rm{g}}}}}{\mathcal{L}^{\nu_{\rm{I}}^{\rm{w}}}}}}} }}} \right)}
     \! d{e_{\rm{w}}\!(m)},\label{yPDF2}
\end{alignat}
where $\int_{\mathbb{C} } {\cdot}$ represents an integral over the entire complex plane.
However, it is difficult to compute the integral in~\eqref{yPDF2} to obtain a closed-form expression.

The DRIS is controlled by the warden Willie, 
so he is aware of the value of $\boldsymbol{\varphi }(m)$ in $h^{\rm w}_{\rm D}(m)$
at the $m$-th sampling time.
Consequently, the conditional PDF of $e_{\rm w}(m)$ can be simplified from~\eqref{ewPDF} to
\begin{alignat}{1}
    \nonumber
    & {{f_{E_{\rm w}}} \!\left( {\left. {{e_{\rm{w}}\!(m)}} \right|{{\mathcal{H}_1},{h_{\rm{D}}^{\rm{w}}}}(m) }   \right) } 
      \sim \mathcal{CN}\!\left( {0,  {\delta^{2}_{\rm e}(m)} } \right) \\
      \nonumber
     &   \;\;\;\;\;\; \; =\frac{1}{{\pi P_0{{\left| \!{{\frac{{h^{\rm{w}}_{{\rm{d}}} } }{{{\mathcal{L}^{\frac{{{\nu^{\rm{w}}_{{\rm{d}}}}}}{2}}}}}}}
     + {{\frac{{h^{\rm{w}}_{{\rm{D}}} (m)} }{{  {\mathcal{L}^{\frac{{{\nu _{{\rm{g}}}}}}{2}}} {\mathcal{L}^{\frac{{{\nu^{\rm{w}}_{{\rm{I}}}}}}{2}}} } }}} \!\right|}^2}}} \\
     & \;\;\;\;\;\;\;\;\;\;\;\;\;\;\;\;\;\;\;\;\; {\exp\!\!\left(\! { - \frac{{ {{\left| {{e_{\rm{w}}}(m)} \right|}^2}}}{{{P_0{{\left| \!{{\frac{{h^{\rm{w}}_{{\rm{d}}} } }{{{\mathcal{L}^{\frac{{{\nu^{\rm{w}}_{{\rm{d}}}}}}{2}}}}}}}
     +{{\frac{{h^{\rm{w}}_{{\rm{D}}} (m)} }{{  {\mathcal{L}^{\frac{{{\nu _{{\rm{g}}}}}}{2}}} {\mathcal{L}^{\frac{{{\nu^{\rm{w}}_{{\rm{I}}}}}}{2}}} } }}}\! \right|}^2}}}}}  \!\right) }.
     \label{ewPDFGua}
\end{alignat}

Conditioned on the fact that the sum of independent complex Gaussian random variables 
is also complex Gaussian, the conditional PDF of the received sample $y_{\rm w}(m)$ at Willie 
simplifies to 
\begin{alignat}{1}
    \nonumber
    & {f_{Y_{\rm w}}\!\left({\left. { y_{\rm w}(m)} \right|{{\mathcal{H}_1},{h_{\rm{D}}^{\rm{w}}(m)}} } \right) } 
      \sim \mathcal{CN}\!\left( {0,  {\delta^{2}_{\rm e}(m) + {\delta^{2}_{\rm w}  } } } \right) \\
      \nonumber
     & \;\;\;\;\;\; \; =  \frac{1}{{\pi \!\!\left( \!\!{\delta _{\rm{w}}^2 \!\!+\!\! P_0{{\left| \!{{\frac{{h^{\rm{w}}_{{\rm{d}}} } }{{{\mathcal{L}^{\frac{{{\nu^{\rm{w}}_{{\rm{d}}}}}}{2}}}}}}}
     \!+\!\! {{\frac{{h^{\rm{w}}_{{\rm{D}}} (m)} }{{  {\mathcal{L}^{\frac{{{\nu _{{\rm{g}}}}}}{2}}} \!\!{\mathcal{L}^{\frac{{{\nu^{\rm{w}}_{{\rm{I}}}}}}{2}}} } }}} \!\right|}^2}}  \right)}} \times \\
     & \;\;\;\;\;\;\;\;\;\;\;\;\;\;\;\;\;\;\;\;\; {\exp\!\!\left(\!\! { - \frac{{\left| {{y_{\rm{w}}}\left( m \right)} \right|^2}}{\delta _{\rm{w}}^2 \!\!+\!   P_0{{\left| \!{{\frac{{h^{\rm{w}}_{{\rm{d}}} } }{{{\mathcal{L}^{\frac{{{\nu^{\rm{w}}_{{\rm{d}}}}}}{2}}}}}}}
     \!+\!\! {{\frac{{h^{\rm{w}}_{{\rm{D}}} (m)} }{{  {\mathcal{L}^{\frac{{{\nu _{{\rm{g}}}}}}{2}}} \!\!{\mathcal{L}^{\frac{{{\nu^{\rm{w}}_{{\rm{I}}}}}}{2}}} } }}} \!\!\right|}^2}}} \!\!\right)}.   
     \label{ywPDFGua}
\end{alignat}
On the other hand, the PDF of the $m$-th sample ${ {y}}_{\rm w}(m)$ received at Willie 
under hypothesis $\mathcal{H}_0$ is equal to that of the AWGN:
\begin{equation}
    {f_{Y_{\rm w}}\!\left({\left. { y_{\rm w}(m)} \right|{\mathcal{H}_0}} \right) } 
   \sim \mathcal{CN}\!\left( {0,  {\delta^{2}_{\rm w}  } } \right)  
    =\frac{1}{{\pi \delta _{\rm{w}}^2}}{e^{ - \frac{{\left| {{y_{\rm{w}}} ( m )} \right|^2}}{{\delta _{\rm{w}}^2}}}}.
    \label{ywPDFH0}
\end{equation}

Based on~\eqref{ywPDFGua} and~\eqref{ywPDFH0},
we derive the detection threshold $\varepsilon(m)$ 
for~\eqref{decisionrule1} in Proposition~\ref{Proposition11}.
\textcolor[rgb]{0.00,0.00,1.00}{Under the PDFs given in~\eqref{ywPDFGua} and~\eqref{ywPDFH0},
$\left|{y_{\rm w}(m)}\right|^2$ is a sufficient statistic 
for the variance parameter, 
 and the LRT depends on ${{y_{\rm{w}}}\left( {{m}} \right)}$ 
 only via  ${\left| {{y_{\rm{w}}}\left( {{m}} \right)} \right|}^2$.}

\begin{proposition}
    \label{Proposition11}
For the detection in~\eqref{decisionrulexx} and~\eqref{decisionrule} performed by Willie, 
the optimal detection thresholds $\varepsilon(m)$ ($m=1,\cdots, M$) based 
on the  LRT are given by
\begin{alignat}{1}
    \nonumber
    \varepsilon(m) &=\frac{{ \left(\!{\delta _{\rm{w}}^2 \!+\! P_0{{\left| \!{{\frac{{h^{\rm{w}}_{{\rm{d}}} } }{{{\mathcal{L}^{\frac{{{\nu^{\rm{w}}_{{\rm{d}}}}}}{2}}}}}}}
    + \! {{\frac{{h^{\rm{w}}_{{\rm{D}}} (m)} }{{  {\mathcal{L}^{\frac{{{\nu _{{\rm{g}}}}}}{2}}} {\mathcal{L}^{\frac{{{\nu^{\rm{w}}_{{\rm{I}}}}}}{2}}} } }}} \!\right|}^2}} \right)\!
    \delta _{\rm{w}}^2}}{{P_0{\left| \!{{\frac{{h^{\rm{w}}_{{\rm{d}}} } }{{{\mathcal{L}^{\frac{{{\nu^{\rm{w}}_{{\rm{d}}}}}}{2}}}}}}}
    +\! {{\frac{{h^{\rm{w}}_{{\rm{D}}} (m)} }{{  {\mathcal{L}^{\frac{{{\nu _{{\rm{g}}}}}}{2}}} {\mathcal{L}^{\frac{{{\nu^{\rm{w}}_{{\rm{I}}}}}}{2}}} } }}} \!\right|}^2}} \times\\
    &\;\;\;\;\;  \left(\!{\ln {\!\!\left(\!{\delta _{\rm{w}}^2 \!+\! P_0{{\left| \!{{\frac{{h^{\rm{w}}_{{\rm{d}}} } }{{{\mathcal{L}^{\frac{{{\nu^{\rm{w}}_{{\rm{d}}}}}}{2}}}}}}}
    + \! {{\frac{{h^{\rm{w}}_{{\rm{D}}} (m)} }{{  {\mathcal{L}^{\frac{{{\nu _{{\rm{g}}}}}}{2}}} {\mathcal{L}^{\frac{{{\nu^{\rm{w}}_{{\rm{I}}}}}}{2}}} } }}} \!\right|}^2}}\right)} - \ln {\delta^2_{\rm{w}}}}\!\right) .
    \label{OptimalLRT}
\end{alignat}
\end{proposition}

\begin{IEEEproof}
See Appendix~\ref{AppendixA1}
\end{IEEEproof}

\textcolor[rgb]{0.00,0.00,1.00}{According to the Bayes detection criterion, the LRT used in Proposition~\ref{Proposition11}
can equivalently minimize the overall detection error probability.
Assuming equal priors, the probabilities of ${{\mathcal{H}}_0}$ and ${{\mathcal{H}}_1}$
are the same.
In this work, we will define the overall detection error rate as
half the sum of the FAR and MDR.} 
According to Proposition~\ref{Proposition11}, the introduction of the DRIS directly affects 
the detection thresholds $\varepsilon(m)$ for the detection rule in~\eqref{decisionrule1}. 
\textcolor[rgb]{0.00,0.00,1.00}{In the presence of a DRIS, 
 $\varepsilon(m)$ varies with $m$.}
Without a DRIS,  the detection thresholds in~\eqref{OptimalLRT}  reduce to
the following fixed value:
\begin{equation}
    \varepsilon =\frac{{ \left(\!{\delta _{\rm{w}}^2 + \frac{{{P_0}{{\left| {h_{\rm{d}}^{\rm{w}}} \right|}^2}}}{{{\mathcal{L}^{\nu_{\rm{d}}^{\rm{w}}}}}}}\!\right)\!
    \delta _{\rm{w}}^2}}{{{{ \frac{{{P_0}{{\left| {h_{\rm{d}}^{\rm{w}}} \right|}^2}}}{{{\mathcal{L}^{\nu_{\rm{d}}^{\rm{w}}}}}}}} }} 
      \!\left(\!{\ln {\!\!\left(\!{\delta _{\rm{w}}^2 + \frac{{{P_0}{{\left| {h_{\rm{d}}^{\rm{w}}} \right|}^2}}}{{{\mathcal{L}^{\nu_{\rm{d}}^{\rm{w}}}}}}}\right)} - \ln {\delta^2_{\rm{w}}}}\right) .
    \label{OptimalLRTred}
\end{equation}

\textcolor[rgb]{0.00,0.00,1.00}{It follows from~\eqref{ywPDFGua} that
the PDF of the total power of ${\boldsymbol{y}}_{\rm w}$ (i.e., $\left\|{\boldsymbol{y}}_{\rm w}\right\|^2$)
is a weighted
sum of exponentials with distinct rates (i.e., a hypoexponential).
Consequently, the LRT 
yields a transcendental equation that, 
in general, admits no closed-form solution.
 Hence, a closed-form optimal detection threshold for the test statistic  $\left\|{\boldsymbol{y}}_{\rm w}\right\|^2$
is generally unavailable. Therefore, in the presence of a DRIS, 
it is difficult for Willie  to directly
perform detection based on $\left\|{\boldsymbol{y}}_{\rm w}\right\|^2$.}

\subsection{Error Probability Analysis of Willie's Detection}\label{BERDec}
In this section, 
we quantify the impact of the time-varying DRIS-based FPJ on the decision-making process employed by the warden Willie.
In particular,  the theoretical FAR $p_{\rm F}$ 
and the MDR $p_{\rm M}$  are derived based on the detection thresholds $\varepsilon(m)$ given by  Proposition~\ref{Proposition11}.
We provide closed-form expressions for $p_{\rm F}$ and $p_{\rm M}$ in Theorem~\ref{Theorem11}.
\newtheorem{theorem}{Theorem} 
\begin{theorem}
 \label{Theorem11}
The FAR and MDR can be expressed as
\begin{alignat}{1}
    \nonumber
    &p_{\rm{F}}  =   {\mathbb{P}}\! \left( {\boldsymbol{y}}_{\rm{w}} \in {\mathcal{S}}|{{\mathcal{H}}_0} \right) \\
    &=  {\sum\limits_{T = N}^{M} {\sum\limits_{{i_1} <  \cdots  < {i_T}} \!  
    \!\!\left({ {\mathop \prod \limits_{j = {i_1}}^{{i_T}} \!e^{\frac{{ - \varepsilon(j)}}{{\delta _{\rm{w}}^2}}}} 
    \!\!\mathop \prod \limits_{i \ne {i_1} \ne  \cdots  \ne {i_T}}\!\!\!\! 
    \left(\! {1 - e^{- \frac{{\varepsilon (i)}}{{\delta _{\rm{w}}^2}}}} \!\right)}\!\!\right) } } , \label{FARExpre1}
\end{alignat}
and
\begin{alignat}{1}
    \nonumber
    &p_{\rm{M}}  = {\mathbb{P}}\! \left( {\boldsymbol{y}}_{\rm{w}} \notin  {\mathcal{S}}|{{\mathcal{H}}_1} \right) \\
    \nonumber
    & = \frac{ \mathop \Pi \limits_{m = 1}^M 
    {\!\!\left(\!\!1-\exp\!\!\left(\!{-\frac{\varepsilon(m)}{{{\delta _{\rm{w}}^2 + \frac{{{P_0}{{\left| {h_{\rm{d}}^{\rm{w}}} \right|}^2}}}{{{\mathcal{L}^{\nu_{\rm{d}}^{\rm{w}}}}}} + 
    \frac{{{P_0}{{\left| {h_{\rm{D}}^{\rm{w}}(m)} \right|}^2}}}{{{\mathcal{L}^{{\nu_{\rm{g}}}}}{\mathcal{L}^{\nu_{\rm{I}}^{\rm{w}}}}}}}}}} \!\right) \!\!\right)} }{2^M} + \\
    \nonumber
    & \sum\limits_{T \!=\! 1}^{N-1} \!  {{\sum\limits_{{i_1} <  \cdots  < {i_T}} \!\! 
    \left(\!\!{\mathop \prod \limits_{j = {i_1}}^{{i_T}} \! {\exp\!\!\left(\!{\! \frac{{ - \varepsilon(j)}}{{{{\delta _{\rm{w}}^2 \!+\! P_0{{\left| \!{{\frac{{h^{\rm{w}}_{{\rm{d}}} } }{{{\mathcal{L}^{\frac{{{\nu^{\rm{w}}_{{\rm{d}}}}}}{2}}}}}}}
    + \! {{\frac{{h^{\rm{w}}_{{\rm{D}}} (j)} }{{  {\mathcal{L}^{\frac{{{\nu _{{\rm{g}}}}}}{2}}} {\mathcal{L}^{\frac{{{\nu^{\rm{w}}_{{\rm{I}}}}}}{2}}} } }}} \!\right|}^2}}}}}}\!\right)}   \times}\right. }} \\
    &  \left.\mathop \prod \limits_{i \ne {i_1} \ne  \cdots  \ne {i_T}}\!\!  
    \left(\! {1 - \exp\!\!\left(\!{- \!\frac{{\varepsilon (i)}}{{\delta _{\rm{w}}^2 \!+\! P_0{{\left| \!{{\frac{{h^{\rm{w}}_{{\rm{d}}} } }{{{\mathcal{L}^{\frac{{{\nu^{\rm{w}}_{{\rm{d}}}}}}{2}}}}}}}
    + \! {{\frac{{h^{\rm{w}}_{{\rm{D}}} (i)} }{{  {\mathcal{L}^{\frac{{{\nu _{{\rm{g}}}}}}{2}}} {\mathcal{L}^{\frac{{{\nu^{\rm{w}}_{{\rm{I}}}}}}{2}}} } }}} \!\right|}^2}}}}\!\right)} \!\!\right)\!\!\!\right). \label{MDRExpre}
\end{alignat} 
\end{theorem}

\begin{IEEEproof}
 See Appendix~\ref{AppendixB}
\end{IEEEproof}

Based on Theorem~\ref{Theorem11}, we can 
obtain important properties of the DRIS-based FPJ for covert communications. 
The larger the transmit power $P_0$ at Alice, the smaller the FAR and MDR at Willie. 
In other words, increasing Alice's transmit power does not effectively improve the communication rate between Alice and Bob due to the DRIS-based FPJ.
Instead, it increases the risk of detection by the warden Willie.
Moreover, from~\eqref{FARExpre1} and~\eqref{MDRExpre}, the improvement in FAR and MDR due to the use of the DRIS by Willie is independent of 
the specific values of the phase shifts of the DRIS.
For example,  only 
1-bit quantized DRIS reflection coefficients are sufficient to enhance the detection accuracy at Willie.

\subsection{Communication Performance for Alice and Bob}\label{SNRDec}
The introduction of the DRIS by Willie not only improves  his detection performance
but also decreases quality of the covert transmissions between Alice and Bob.
The time-varying DRIS can effectively jam the transmission between Alice and Bob even 
when Willie experiences a missed detection, since we see from~\eqref{eqSLNR} that
there is an additional DRIS-based ACA interference that degrades Bob's reception performance.
To mathematically characterize the impact of this interference, 
the ergodic SJNR at Bob derived in Theorem~\ref{Theorem2} below.

\begin{theorem}
    \label{Theorem2}
    The ergodic SJNR at Bob converges in distribution to
    \begin{equation}
        {\eta _{\rm b}} =  
        \frac{\frac{ P_0 }{{{\mathcal{L}^{{{{\nu^{\rm{b}}_{{\rm{d}}}}}}}}}}}
        { \frac{P_0  N_{\rm D}{\overline \alpha} }{{{\mathcal{L}^{{{{\nu _{{\rm{g}}}}}}}}}{{\mathcal{L}^{{{{\nu^{\rm{b}}_{{\rm{I}}}}}}}}}} + \delta _{\rm{b}}^2},
        \;{\rm{as}} \; N_{\rm D} \to \infty.
        \label{SNRBobVau}
    \end{equation}
\end{theorem}
   
\begin{IEEEproof}
See Appendix~\ref{AppendixC}.
\end{IEEEproof}
 
We see from Theorem~\ref{Theorem2} that 
increasing the transmit power does not result in an indefinite increase in the SJNR ${\eta _{\rm{b}}}$.
Instead, ${\eta _{\rm{b}}}$ asymptotically converges to a constant value 
$\frac{{\mathcal{L}^{{{{\nu_{{\rm{g}}}}}}}} {\mathcal{L}^{{{{\nu^{\rm{b}}_{{\rm{I}}}}}}}}}{{\mathcal{L}^{{{{\nu^{\rm{b}}_{{\rm{d}}}}}}}}N_{\rm D}{\overline \alpha}}$ 
as $P_0  \to \infty$.
However, based on Theorem~\ref{Theorem11},
the  FAR and MDR at  Willie decrease rapidly 
as the Alice transmit  power $P_0$ increases.
It is worth emphasizing here that the implementation of the DRIS
relies on neither the CSI of the channels between Alice and Bob nor additional jamming power.

\section{Simulation Results and Discussion}\label{ResDis}
In this section, we provide numerical results to demonstrate the impact of  the DRIS on covert communication 
and evaluate the effectiveness of the theoretical analysis in Section~\ref{Analysis}. 
Default settings for the simulation parameters are given below.
As shown in Fig.~\ref{figLoc}, the single-antenna Alice is located at (0m, 0m, 5m),
while the single-antenna Bob is randomly distributed within the ring-shaped region centered at 
(0m, 140m, 0m) with radii between 10m to 20m. 
The warden Willie is located at (0m, 100m, 0m) and monitors
the covert communications between Alice and Bob 
assisted by a DRIS. The DRIS is equipped with 2048 ($N_{{\rm D},h} = 64,N_{{\rm D},v} = 32$) reflective elements and
deployed at (-$d_{\rm{AD}}$ m, 0 m, 5m),
and the distance between Alice and the center of the DRIS i.e., $d_{\rm{AD}}$ is nominally set to 1.5.
The DRIS employs one-bit quantized phase shifts and amplitudes taken from 
$\Psi = \{\frac{\pi}{9},\frac{7\pi}{6}\}$ and 
$\Omega  = {\cal F}\!\left({\Theta}\right) = \{0.8,1\}$~\cite{PGFun},
and the two phase shifts are chosen with equal probability. 
Consequently, $\overline \alpha $ in Proposition~\ref{Proposition1} is calculated to be 0.82.
We assume that the number of samples used for detection by Willie within the channel coherence time 
is $M=5$ and $N=2$ in the detection rule~\eqref{decisionrule1}.
The large-scale LoS and NLoS fading coefficients are defined in Table~\ref{tab1} based on the 3GPP propagation models~\cite{3GPP}, 
and the variance of the noise is $\sigma^2_{\rm c}\!=\!-170\!+\!10\log_{10}\left(BW\right)$ dBm with a transmission bandwidth of 180 kHz.
 
\begin{figure}[!t]
    \centering
    \includegraphics[scale=0.85]{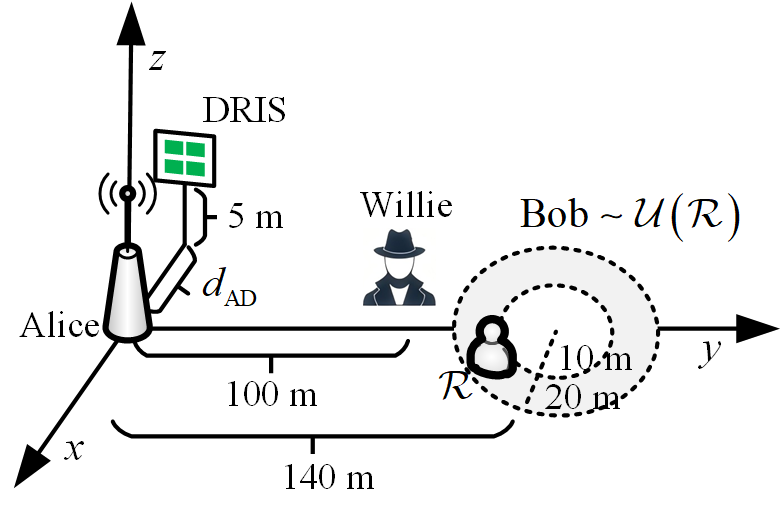}
    \caption{An example of covert communications in the presence of a DRIS, 
    where Bob is randomly located in the ring-shaped region $\mathcal{R}$ 
    centered at (0m, 140m, 0m) with a uniform distributed across radii between 10m to 20m, 
    and Alice and the DIRS are deployed at (0 m, 0 m, 5m) and (-$d_{\rm{AD}}$ m, 0 m, 5m) respectively.}
    \label{figLoc}
\end{figure}

\begin{table}
    \footnotesize
    \centering
    \caption{Wireless Channel Simulation Parameters}
    \label{tab1}
    \begin{threeparttable}
    \begin{tabular}{ c|c }
    \hline
    Large-scale Parameter        &Value\\
    \hline
    LoS fading       & $ 35.6 + 22{\log _{10}}({d}) $ (dB) \\
    \hline
    NLoS fading      &$32.6+36.7{\log _{10}}({d})$ \\
    \hline
    \end{tabular}
    \end{threeparttable}
\end{table} 

\begin{figure}[htbp]
    \centerline{\includegraphics[scale=0.075]{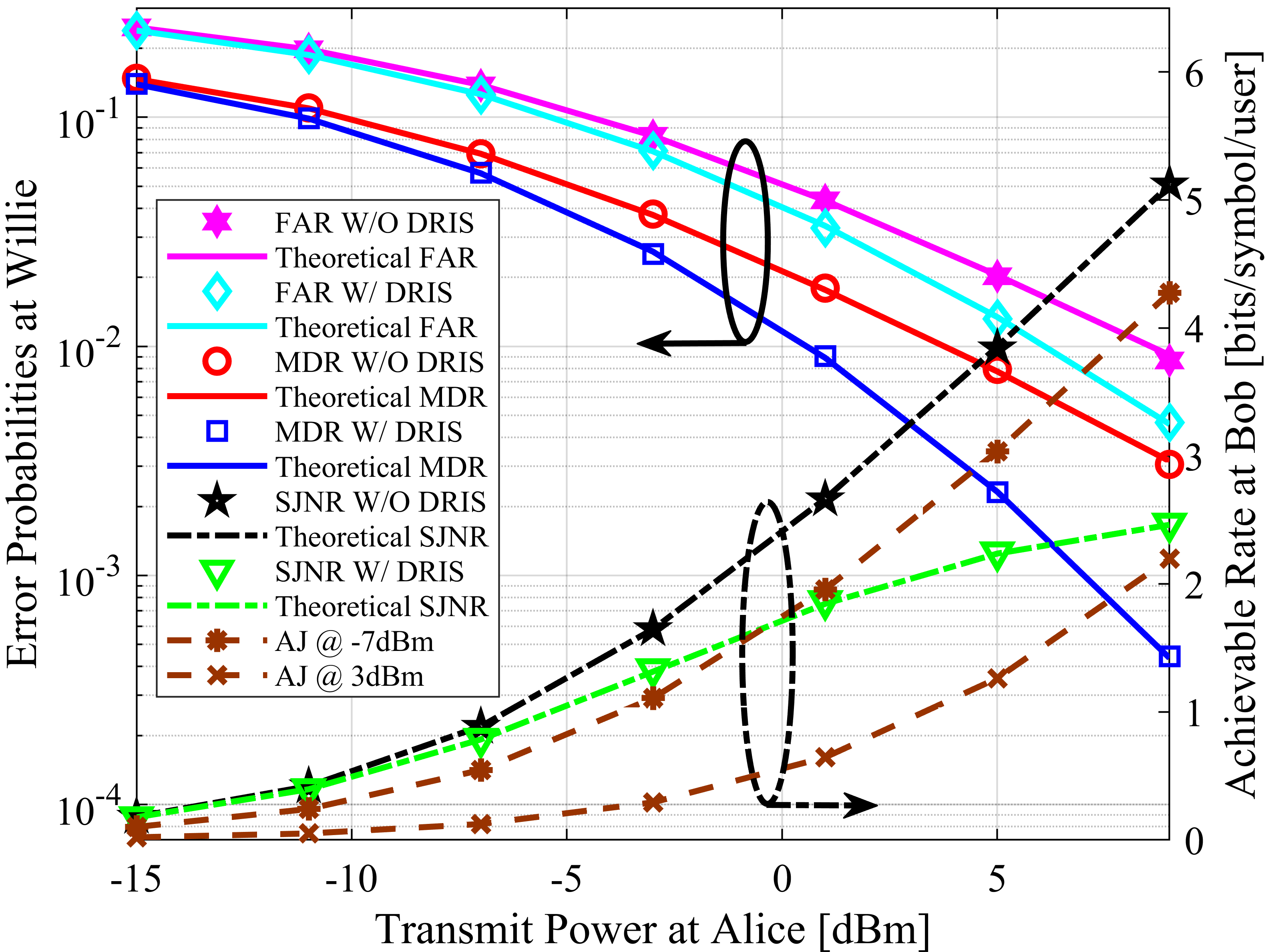}}
    \caption{Relationship between FAR and MDR and transmit power (left y-axis),  
    and that between achievable rate and transmit power (right y-axis).}
    \label{Resfig1}
\end{figure}

\textit{1) Impact of Transmit Power at Alice:} 
Fig.~\ref{Resfig1} illustrates the FAR and MDR at Willie (left y-axis) versus the transmit power at Alice 
obtained from the following benchmarks:
i) the simulated FAR without DRIS (FAR W/O DRIS)~\cite{CCRef1,CCRef11} and ii) the corresponding theoretical FAR without DRIS, 
iii) the simulated FAR with DRIS (FAR W/ DRIS) and iv) the derived theoretical FAR with DRIS in~\eqref{FARExpre1}; 
v) the simulated MDR without DRIS (MDR W/O DRIS)~\cite{CCRef1,CCRef11} and vi) the corresponding theoretical MDR without DRIS;
vii) the simulated MDR with DRIS (MDR W/ DRIS) and viii) the derived theoretical MDR with DRIS in~\eqref{MDRExpre}.
The achievable rates at Bob (right y-axis) versus the transmit power at Alice are also presented in Fig.~\ref{Resfig1}, comparing the following benchmarks:
i) the simulated achievable rates without DRIS (Rate W/O DRIS) and 
ii) the theoretical achievable rates without DRIS;
and iii) the simulated achievable rates with DRIS (Rate W/ DRIS) and iv) the theoretical achievable rates with DRIS based on~\eqref{SNRBobVau}.

As illustrated in Fig.~\ref{Resfig1}, the introduction of the DRIS by Willie significantly improves 
his FAR and MDR.
Meanwhile, the DRIS significantly disrupts covert communications between Alice and Bob by launching FPJ attacks.
As mentioned above, the use of a DRIS by Willie requires neither the CSI of the Alice-Bob channel 
nor additional jamming power.
Fig.~\ref{Resfig1} also verifies the validity of Theorems~\ref{Theorem11} and~\ref{Theorem2}, 
whose theoretical predictions match the results obtained from Monte Carlo simulations.
{\textcolor[rgb]{0.00,0.00,1.00}{An active
jammer (AJ) can also be used by Willie to reduce the SJNR at Bob without requiring Bob's CSI, 
although this approach requires additional jamming power.
The level of covertness diminishes as the ``difference'' between  
${\mathcal{H}}_0 $ and ${\mathcal{H}}_1$ increases, as 
commonly measured by metrics such as the Kullback-Leibler (KL) divergence~\cite{CCRef1,CCRef2}.
The active jamming signals are present under both ${\mathcal{H}}_0 $ and ${\mathcal{H}}_1$, 
and thus the AJ will not significantly decrease the detection error probabilities.
Therefore, we only show the impact of an AJ on the communication performance
metric, i.e., SJNR. Specifically, the AJ continuously broadcasts
jamming signals~\cite{MyWCMag,DIRSTWC, TWCAnti}, 
such as modulated Gaussian waveforms or pseudorandom noise, 
and it is assumed to be deployed at the same location as the DRIS. 
In the simulations below, both low (-7 dBm) and high power (3 dBm) jamming cases are considered.
}}

Both the detection improvement at Willie and the FPJ impact on covert communications
increase with the transmit power at Alice, as shown in Fig.~\ref{Resfig1}.
We see that increasing the transmit power at Alice does not significantly improve covert communication performance due to the DRIS-based FPJ.
Furthermore, as Alice increases her transmit power, the risk of being detected by 
Willie also increases, 
while the covert communications between Alice and Bob suffers increased FPJ.
{\textcolor[rgb]{0.00,0.00,1.00}{However, increasing the transmit power at Alice can effectively mitigate the active jamming impact. 
Moreover, it can be seen that 
the impact of active jamming is positively correlated with the jamming power at the AJ. 
Unfortunately, in covert communications, an AJ deployed by Willie to
jam the communication between Alice and Bob has to remain active at all times, as Willie
cannot know in advance whether Alice and Bob are silent or not. 
This results in significant energy consumption due to continuous jamming, 
particularly when the jamming power is high.}}
In the following discussions, we investigate the impact of different factors 
on covert communications for both low and high transmit powers at Alice.

\textit{2) Impact of Number of DRIS Reflective Elements:} 
 
\begin{figure*}[!t]
    \centering
    \subfloat{
            \includegraphics[scale=0.077]{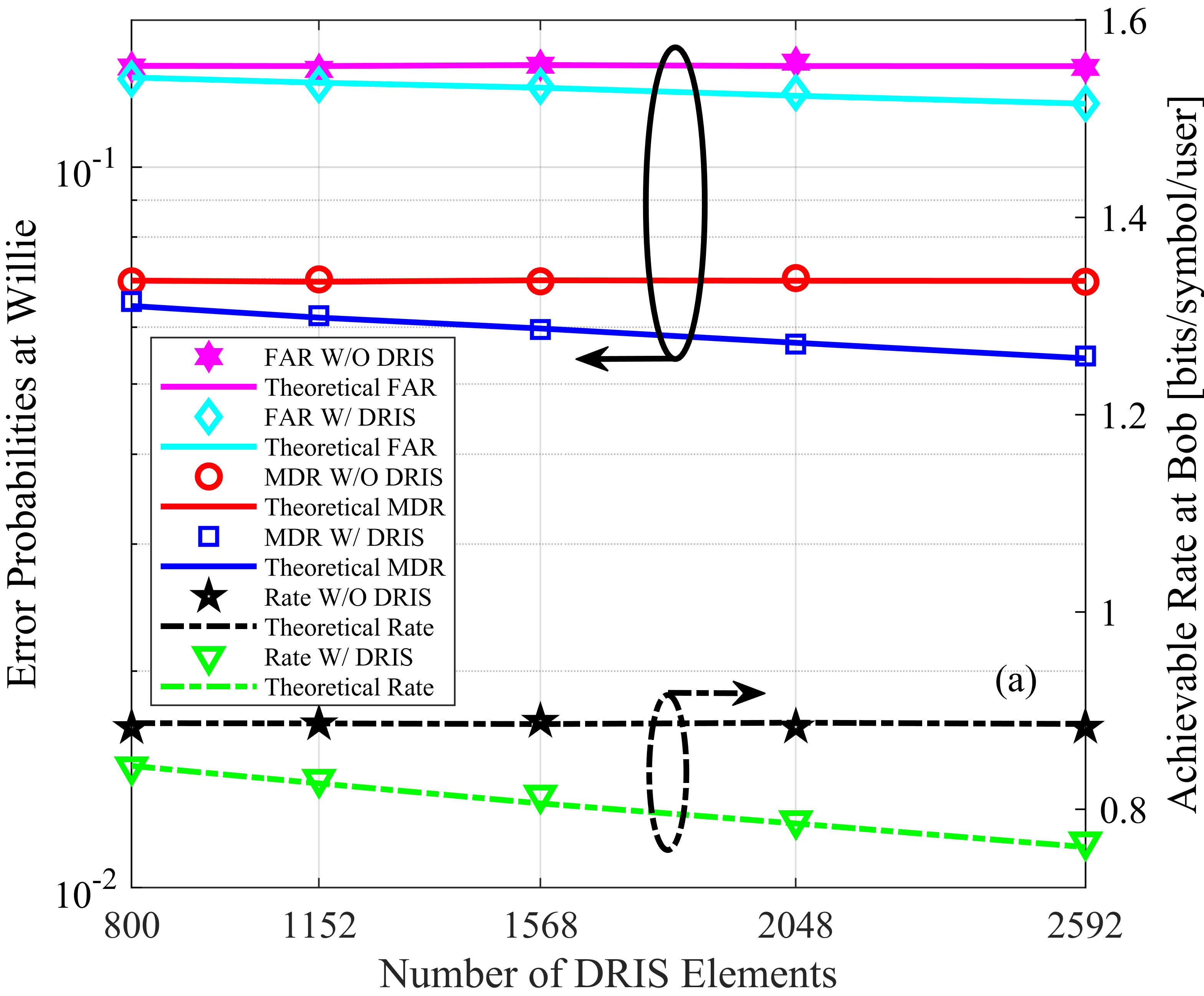}}\hspace{21pt}
    \subfloat{
            \includegraphics[scale=0.077]{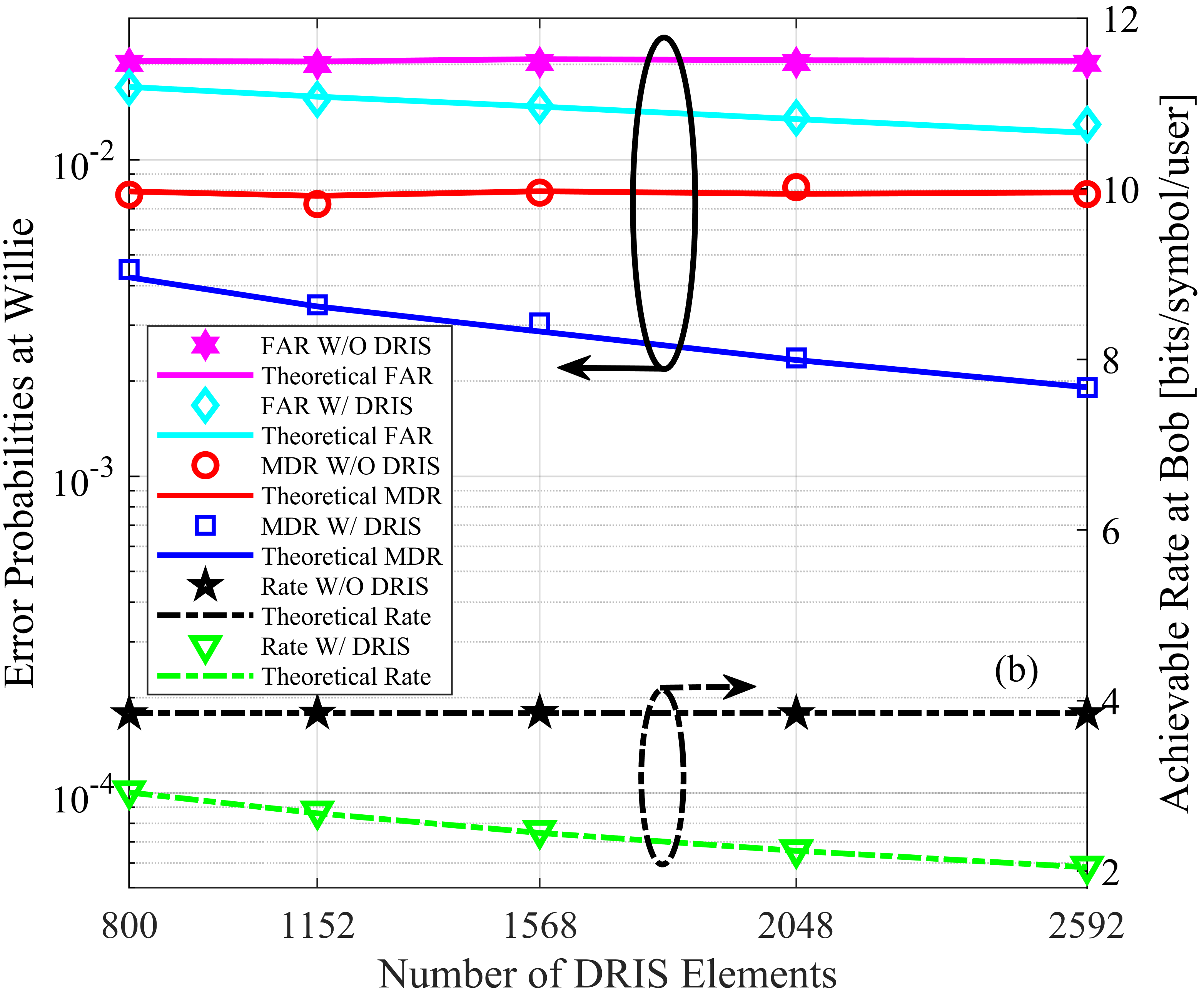}}
   \caption{FAR and MDR vs. the number of DRIS  elements (left y-axis), 
   and achievable rate vs. the number of DRIS elements (right y-axis) at (a) low transmit power (-7 dBm) and (b) high transmit power (5 dBm).}
    \label{ResFigND}
\end{figure*}
Fig.~\ref{ResFigND} illustrates the FAR and MDR at Willie (left y-axis),
as well as the achievable rates at Bob (right y-axis) 
versus the number of DRIS elements for the different benchmarks.
Results obtained for low transmit power (-7 dBm) and high transmit power (5 dBm) 
are plotted in Figs.~\ref{ResFigND} (a) and (b), respectively.
Consistent with Theorem~\ref{Theorem11}, we observe that both the FAR and MDR decrease 
as the number of DRIS elements increases.
Furthermore, we see that the rate of the decrease in the FAR and MDR is also influenced by the transmit power at Alice.
Figs.~\ref{ResFigND} (a) and (b) also show that 
a higher transmit power at Alice leads to a higher detection accuracy at Willie.

Clearly, the impact of the DRIS becomes more pronounced as its number of elements increases. 
Figs.~\ref{ResFigND} (a) and (b) show that the achievable rates at Bob decrease with the number of DRIS reflective 
elements, and Willie can leverage a larger DRIS to reduce his detection error probabilities 
while at the same time degrading the covert transmissions between Alice and Bob.
However, the rate of improvement in the FAR and MDR gradually decrease as the 
size of the DRIS continues to increase.
Nevertheless, again consistent with Theorems~\ref{Theorem11} and~\ref{Theorem2}, 
the detection errors at Willie and the achievable rate at Bob under DRIS-based FPJ degrade to zero 
as the number of DRIS elements approaches infinity.

\begin{figure*}[!t]
    \centering
    \subfloat{
            \includegraphics[scale=0.077]{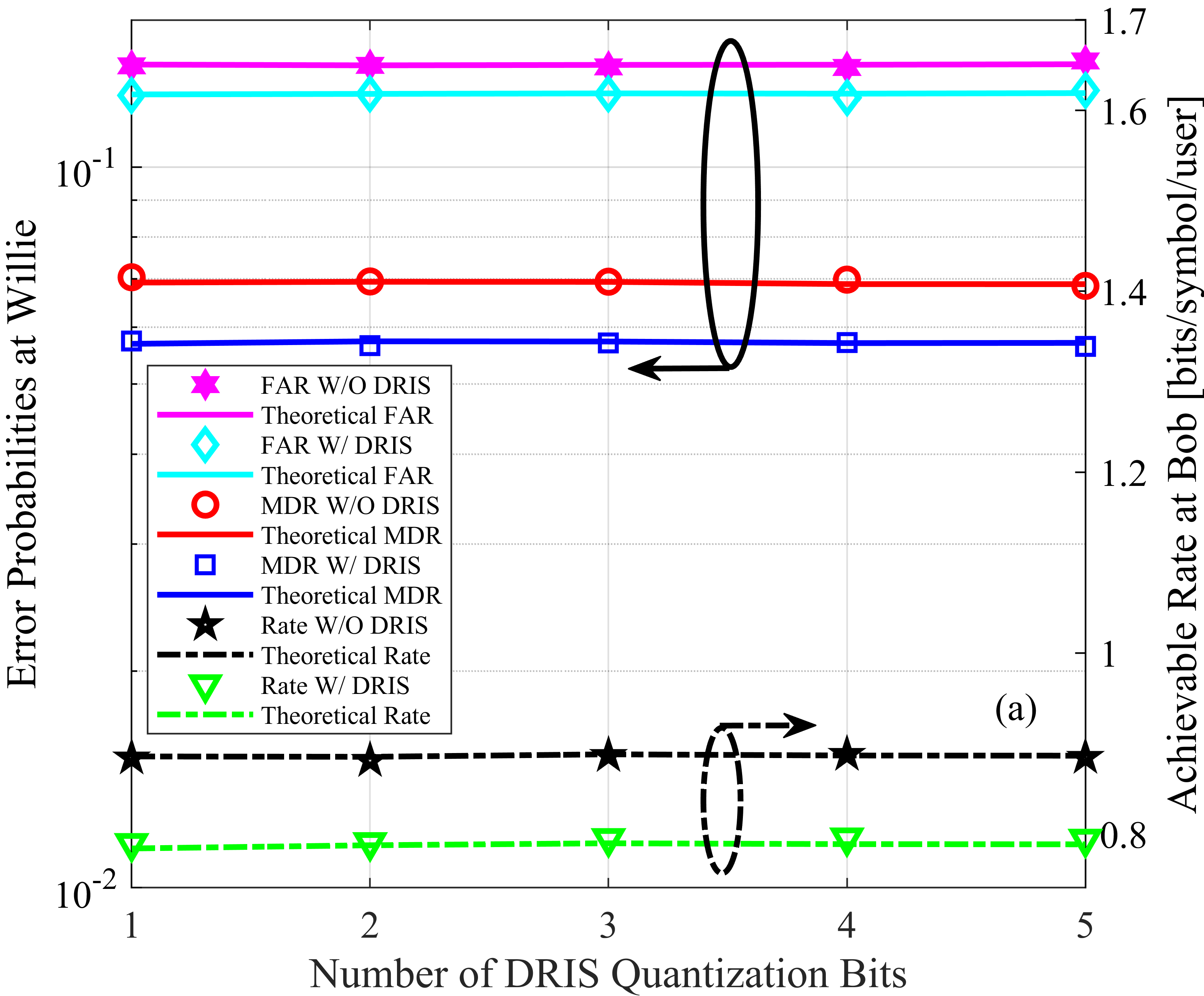}}\hspace{21pt}
    \subfloat{
            \includegraphics[scale=0.077]{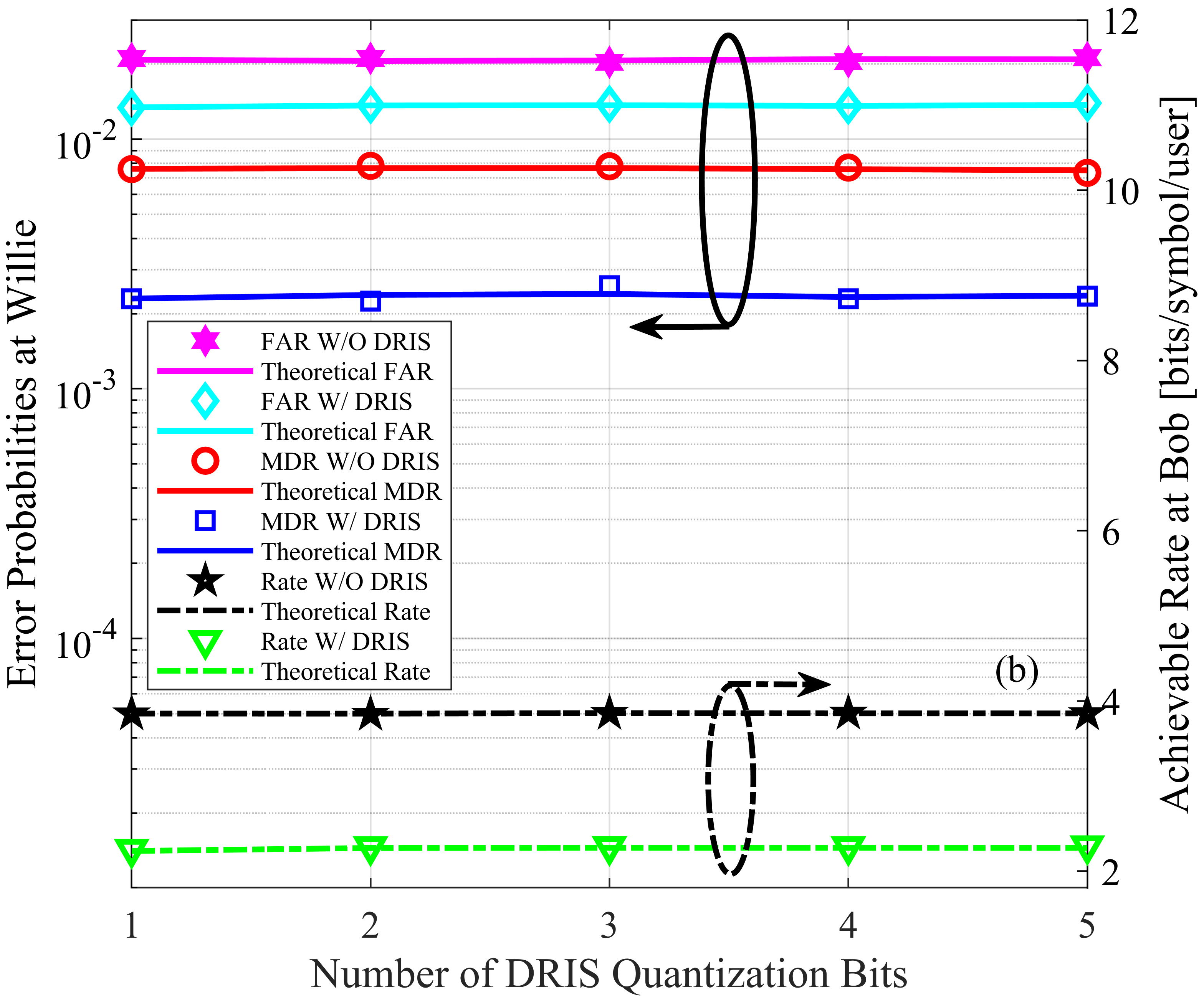}}
   \caption{FAR and MDR vs. the quantization resolution of the DRIS (left y-axis), 
   and achievable rate vs. the quantization resolution of the of DRIS (right y-axis) at (a) low transmit power (-7 dBm) and (b) high transmit power power (5 dBm).}
    \label{ResFigBits}
\end{figure*}

\textit{3) Impact of DRIS Reflection Coefficient Quantization:}  In Fig.~\ref{ResFigBits},
the influence of quantization of the DRIS responses is investigated.
A $b$-bit quantized DRIS with $2^b$ phase shift values is modeled based on~\cite{PGFun1}, 
where the phase shift values are denoted as $\Psi =\{\frac{-\pi}{2}, -\pi+\frac{\pi}{2^{b-1}},\cdots,\frac{3\pi}{2}-\frac{\pi}{2^{b-1}}\}$.
Furthermore, based on~\cite{PGFun1},
the time-varying phase shift and amplitude of the $n$-th DRIS element ($n=1,2,\cdots,N_{\rm D}$) are modeled by
\begin{alignat}{1}
    \nonumber
    \beta_n(t) & = {\mathcal F}\left({\varphi}_{n}(t)\right) \\
    & = (1 - \alpha_{\rm{min}})\!\left(\!\frac{\sin\!\left({\varphi}_r(t) - \phi\right) + 1}{2}\!\right)^{\mu} \!+ \alpha_{\rm{min}},
    \label{PAFun}
\end{alignat}
where ${\varphi}_{n}(t) \in \Psi$ denotes the phase shift of the $n$-th DRIS element, 
$\alpha_{\rm{min}} = \min\{\Omega\}$ represents the minimum DRIS amplitude, 
and $\mu$ and $\phi$ are constants determined by the specific RIS implementation.
According to~\cite{PGFun1}, we set $\alpha_{\rm{min}} = 0.8$, ${\mu} = 1.6$, and $\phi =0$.

In most RIS investigations, increasing the number of quantization bits typically results in a more significant impact.
However, Fig.~\ref{ResFigBits} presents a surprising result, 
showing that increasing the quantization resolution of the DRIS has only a minimal impact on the  FAR and MDR at Willie 
and the achievable rate at Bob.
This is because, according to Proposition~\ref{Proposition1}, 
the statistics of the cascaded DRIS-based channels 
are not related to the specific choice of the DRIS reflection coefficients, 
but rather depend on the expectation of $\left|\beta_{n}(t)\right|^2, n = 1,\cdots, N_{\rm D}$, 
i.e.,
$\overline \alpha =  \mathbb{E} \!\left[\left|\beta_{n}(t) \right|^2\right]= \sum\nolimits_{i = 1}^{{2^b}} {{{\rm P}_i}\alpha _i^2}$.
However, based on~\eqref{PAFun}, increasing the number of DRIS  quantization bits
merely provides more available amplitude values for the DRIS
without significantly increasing the value of $\overline \alpha$.
As a result, a DRIS with higher quantization precision does not lead to a more substantial impact on covert communications. 
Based on the results plotted in Figs.~\ref{ResFigND} and~\ref{ResFigBits}, 
a 1-bit DRIS with a large number of reflective elements 
is sufficient to enhance the detection accuracy at the warden Willie and degrade the covert transmissions between Alice and Bob.
\begin{figure*}[!t]
    \centering
    \subfloat{
            \includegraphics[scale=0.077]{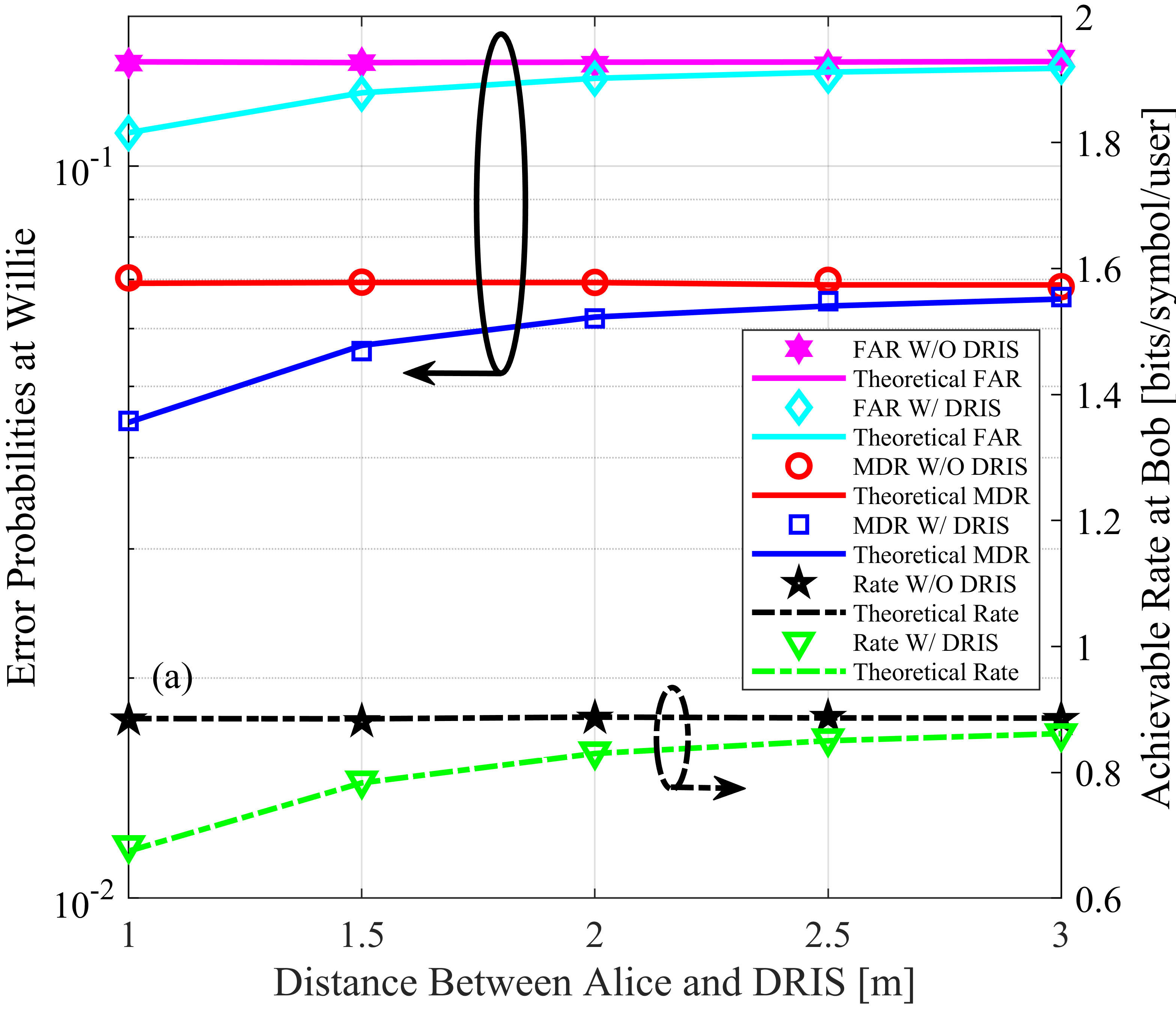}}\hspace{21pt}
    \subfloat{
            \includegraphics[scale=0.077]{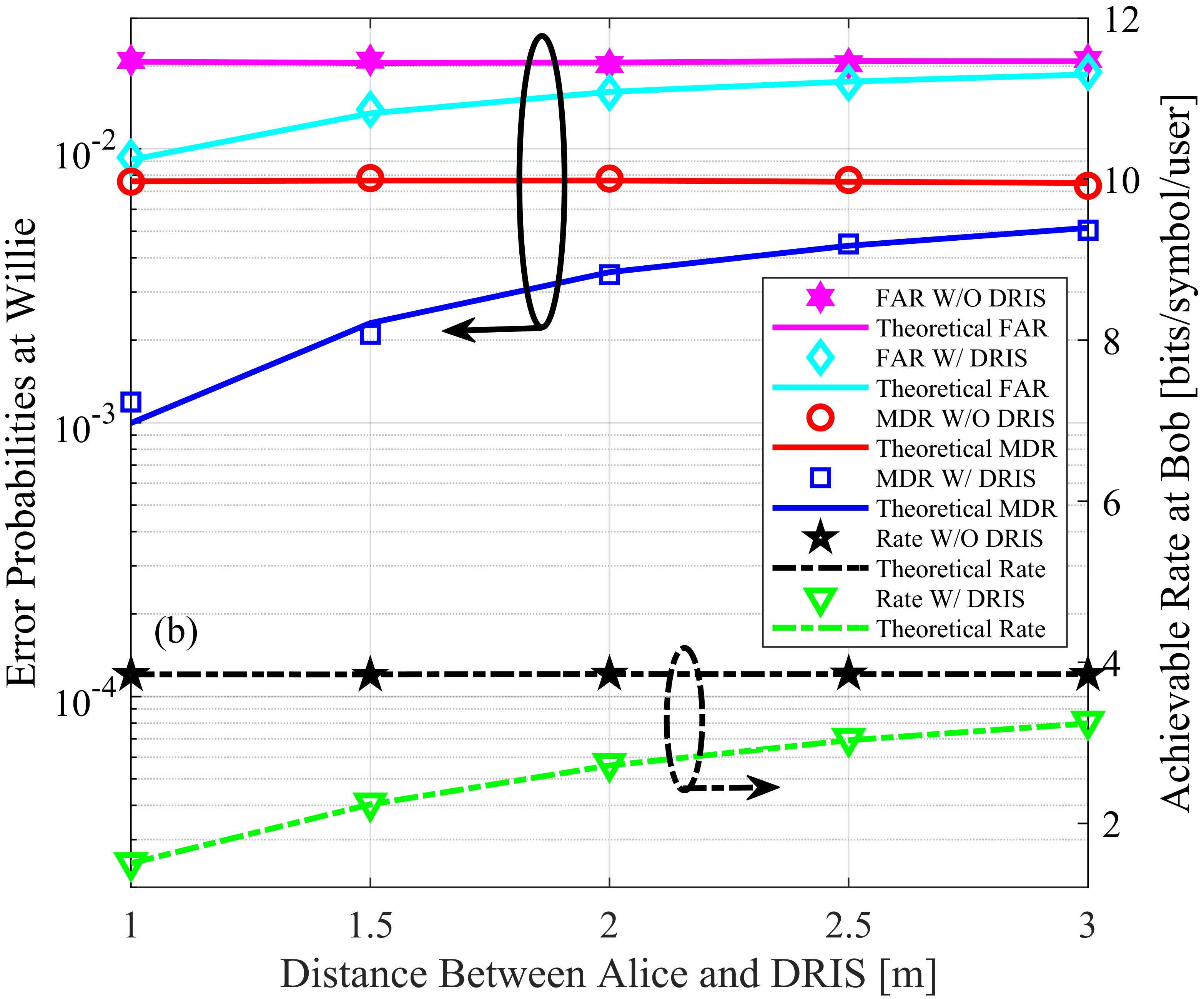}}
   \caption{FAR and MDR vs. the distance between Alice and DRIS (left y-axis), 
   and achievable rate vs. the distance between Alice and DRIS (right y-axis) at (a) low transmit power (-7 dBm) and (b) high transmit power (5 dBm).}
    \label{ResFigdD}
\end{figure*}

\textit{4) Impact of Distance Between Alice and DRIS:}
According to Theorems~\ref{Theorem11} and~\ref{Theorem2}, the impact of DRIS on covert communications 
is related to the large-scale fading of the cascaded DRIS-based channels $h^{\rm w}_{\rm D}(t)$ and $h^{\rm b}_{\rm D}(t)$.
Fig.~\ref{ResFigdD} shows the FAR and MDR at Willie 
and the achievable rate at Bob as functions of the distance between Alice and DRIS $d_{\rm{AD}}$
for low and high transmit power levels, i.e.,  -7 dBm and 5 dBm transmit powers.
We see that the impact of the DRIS on both detection performance at Willie and the achievable rate at Bob 
diminishes as $d_{\rm{AD}}$ increases.
In other words, if the warden Willie aims to closely monitor covert communications between Alice and Bob 
and maximize tthe impact of the DRIS-based FPJ attacks, the DRIS should be deployed as close to Alice as possible.

\begin{figure*}[!t]
    \centering
    \subfloat{
            \includegraphics[scale=0.077]{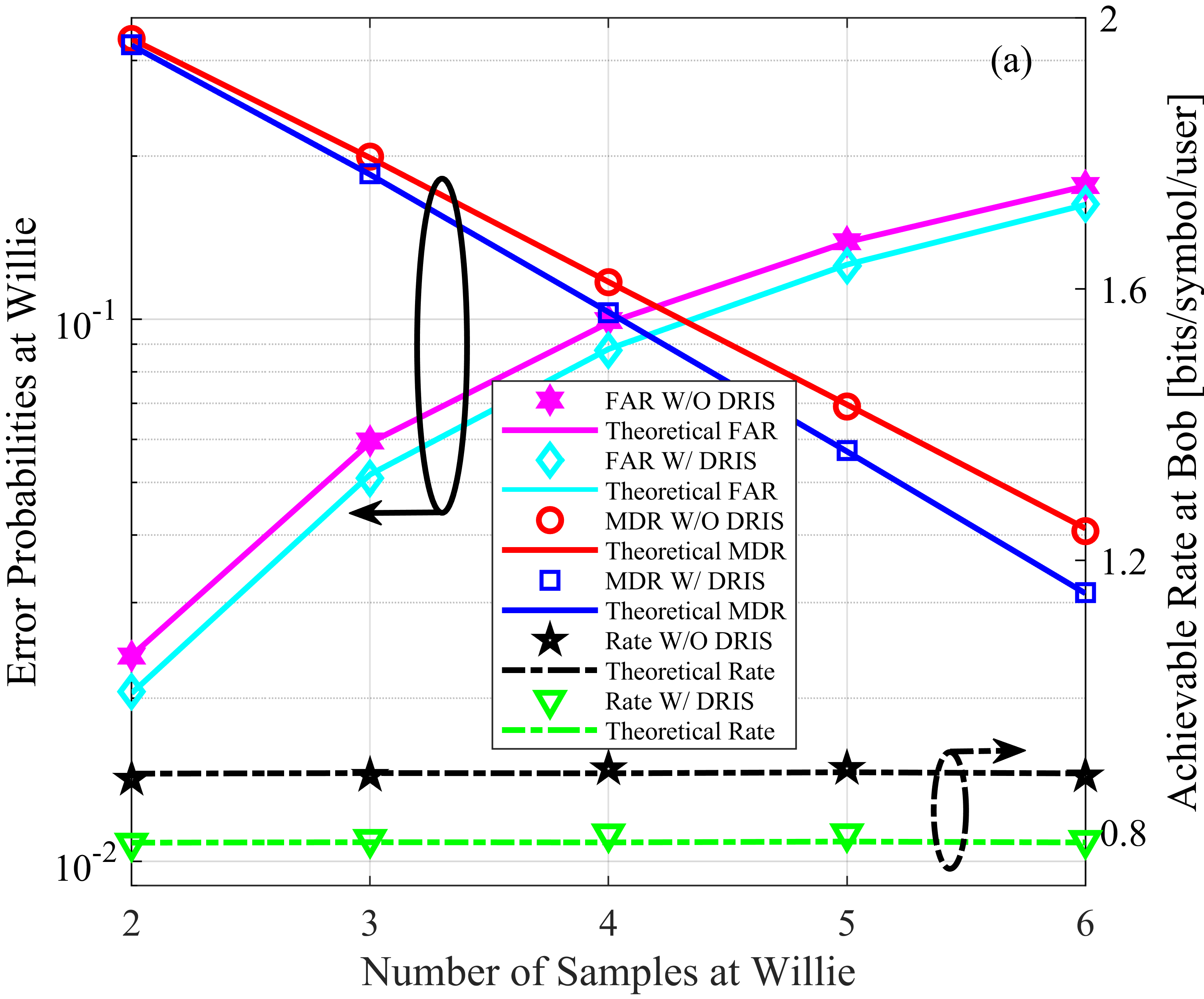}}\hspace{21pt}
    \subfloat{
            \includegraphics[scale=0.077]{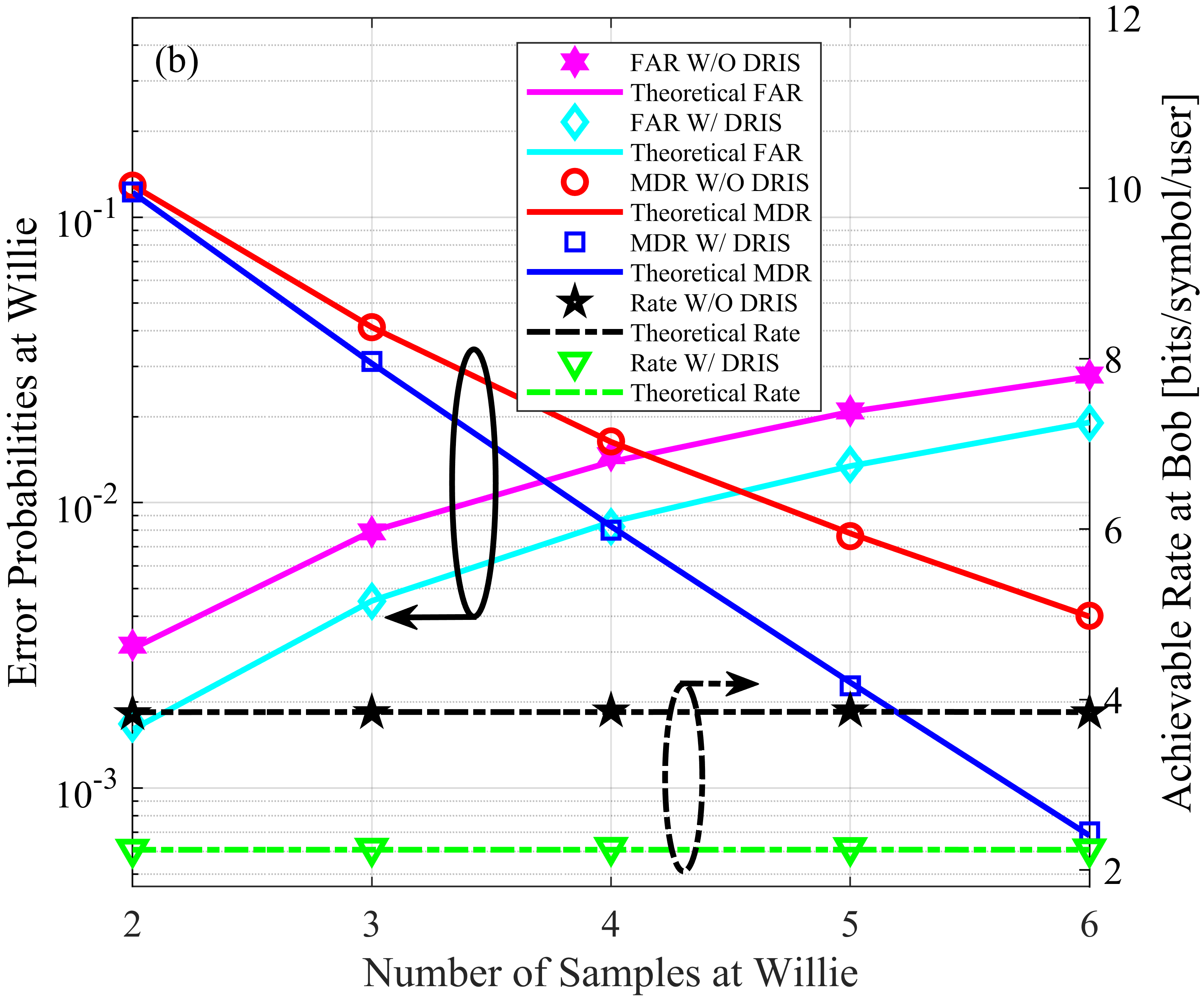}}
   \caption{FAR and MDR vs. the number of Willie's detections (left y-axis), 
   and achievable rate vs. the number of Willie's detections (right y-axis) at (a) low transmit power (-7 dBm) and (b) high transmit power (5 dBm).}
    \label{ResFigTimes}
\end{figure*}

\textit{5) Impact of Number of Detection Samples:}
Fig.~\ref{ResFigTimes} illustrates the relationship between the number of samples used for detection by Willie and achievable rate at Bob.
While the number of samples has no impact on Bob's performance, it significantly
impacts Willies's detection error probabilities, regardless of whether the DRIS is implemented.

We see that the MDR decreases monotonically with the number of Willie's detection samples, while the FAR increases.
The results of Theorem~\ref{Theorem11}
suggest that a trade-off between the FAR and MDR can be achieved by appropriately setting $N$. 
For example, when  Willie makes six detections within the channel coherence time, the parameter
$N$ in the detection rule can be changed from 2 to 3 to achieve a balance between the FAR and MDR.

\section{Conclusions}\label{Conclus} 
In this paper, we proposed a novel approach to help a warden Willie improve his ability to detect the presence of covert communications between Alice and Bob, 
 while simultaneously jamming such communications. 
 The proposed DRIS-based approach requires neither CSI for 
 the Alice-Bob channel nor active jamming power.
The following conclusions can be drawn from the theoretical analysis and numerical results.
\begin{enumerate}
    \item A DRIS with random and time-varying reflection coefficients introduces FPJ in covert communications.
    If the detection rule performed by the warden Willie takes the time-varying DRIS-based FPJ into account, 
    the DRIS not only reduces the FAR and MDR at Willie but also significantly disrupts the covert transmissions between Alice and Bob, 
    even when Willie experiences a missed detection.
    \item Increasing the transmit power at Alice does not significantly improve the communication performance between Alice and Bob 
    due to the DRIS-based FPJ. 
    Moreover, higher transmit power increases the detection accuracy at Willie.
    In addition, the covert communications between Alice and Bob experience more severe DRIS-based FPJ.
    \item A 1-bit DRIS with a large number of reflective elements is sufficient to significantly improve the detection accuracy at the warden Willie and degrade the covert communication performance between Alice and Bob. 
    Moreover, the warden Willie should deploy the DRIS as close to Alice as possible to effectively monitor the communications between Alice and Bob
     and maximize the impact of the DRIS-based FPJ attacks.
\end{enumerate}

Our work demonstrates that illegitimate RISs can pose a significant threat to covert communications, 
even without relying on either channel knowledge or additional jamming power, 
highlighting a critical area of concern.
{\textcolor[rgb]{0.00,0.00,1.00}{Therefore, it is necessary 
that covert communication systems
investigate potential countermeasures, such as signal classification or anomaly detection techniques based on artificial intelligence (AI)~\cite{AIzhang}, 
or adaptive beamforming.}}

\begin{appendices}
\section{Proof of Proposition~\ref{Proposition1}}\label{AppendixA}
    Based on the definition of the cascaded DRIS-based channel between Alice and Willie expressed in~\eqref{hDEx},
    the cascaded DRIS-based channel ${ h}^{\rm w}_{\rm D}(t)$ can be written as
    \begin{alignat}{1}
        \nonumber
        \frac{{ h}^{\rm w}_{\rm D}(t)}{{{ {\mathcal{L}^{\frac{\nu^{\rm w}_{\rm g}}{2}} } 
        {\mathcal{L}^{\frac{\nu_{\rm I}^{ \rm{w}}}{2}}} } }} = 
        &\sqrt {\frac{{{\varepsilon_{\rm g}}}}{{\left({{\varepsilon_{\rm g}}\! + \!1}\right){{{ {\mathcal{L}^{\nu^{\rm w}_{\rm g}} } 
        {\mathcal{L}^{\nu_{\rm I}^{ \rm{w}}}} } }}}}} { {\boldsymbol{g}}}^{{\rm{LOS}}} { {{\boldsymbol{h}}}^{\rm w}_{{\rm I}}} \!\odot\! {\boldsymbol{\varphi}}(t)  \\
        &+ \sqrt {\frac{  1 }{{\left({{\varepsilon_{\rm g}}\! + \!1}\right){{{ {\mathcal{L}^{\nu^{\rm w}_{\rm g}} } 
        {\mathcal{L}^{\nu_{\rm I}^{ \rm{w}}}} } }}}}} { {\boldsymbol{g}}}^{{\rm{NLOS}}}{ {{\boldsymbol{h}}}^{\rm w}_{{\rm I}}} \!\odot\! {\boldsymbol{\varphi}}(t),
        \label{RewriHACAele18}
    \end{alignat}
where  $\odot$ denotes the Hadamard product.

According to the definition of $\boldsymbol{h}^{\rm w}_{\rm I}$ in~\eqref{Hdkeq} and  the definition that
${\boldsymbol \varphi }(t) = \left[\beta_1(t)e^{\varphi_1(t)}, \cdots,\beta_{N_{\rm D}}(t)e^{\varphi_{N_{\rm D}}(t)}   \right]$,  
we rewrite~\eqref{RewriHACAele18} as
    \begin{alignat}{1}
        \nonumber
        &\frac{{ h}^{\rm w}_{\rm D}(t)}{{{ {\mathcal{L}^{\frac{\nu^{\rm w}_{\rm g}}{2}} } 
        {\mathcal{L}^{\frac{\nu_{\rm I}^{ \rm{w}}}{2}}} } }} = \\ \nonumber
        &\;\;\; \sqrt {\frac{{{\varepsilon_{\rm g}} {{ {\mathcal{L}^{\nu^{\rm w}_{\rm g}} } 
        {\mathcal{L}^{\nu_{\rm I}^{ \rm{w}}}} } }}}{\left({{\varepsilon_{\rm g}}\! + \!1}\right){{{ {\mathcal{L}^{\nu^{\rm w}_{\rm g}} } 
        {\mathcal{L}^{\nu_{\rm I}^{ \rm{w}}}} } }}}}
        \sum\limits_{r = 1}^{{N_{\rm{D}}}} \!\left({\!\left[{ {\boldsymbol{g}}}^{{\rm{LOS}}}\right]_{r}\!
        \left[{ {{\boldsymbol{h}}}^{\rm w}_{{\rm I}}} \right]_{r} \!{\beta_r}(t){e^{j{\varphi_r}(t)}}}\!\right) + \\
        &\sqrt {\frac{1}{{\left({{\varepsilon_{\rm g}}\! + \!1}\right){{{ {\mathcal{L}^{\nu^{\rm w}_{\rm g}} } 
        {\mathcal{L}^{\nu_{\rm I}^{ \rm{w}}}} } }}}}} \sum\limits_{r = 1}^{{N_{\rm{D}}}} 
        \left({\!\left[{ {\boldsymbol{g}}}^{{\rm{NLOS}}}\right]_{r}\!
        \left[{  {{\boldsymbol{h}}}^{\rm w}_{{\rm I}}} \right]_{r} \!
        {\beta_r}(t){e^{j{\varphi_r}(t)}}}\!\right).
        \label{Rewri18HACAele1}  
    \end{alignat}
The expectations of the variables in~\eqref{Rewri18HACAele1} are calculated as
\begin{alignat}{1}
 &\mathbb{E} \!\left[\!{{\left[{ {\boldsymbol{g}}}^{{\rm{LOS}}}\right]_{r}\!
 \left[{ {{\boldsymbol{h}}}^{\rm w}_{{\rm I}}} \right]_{r} \!{\beta_r}(t){e^{j{\varphi_r}(t)}}} }\!\right] = 0,\label{LoSExp}\\
 &\mathbb{E} \!\left[ \! \left[{ {\boldsymbol{g}}}^{{\rm{NLOS}}}\right]_{r}\!
 \left[{  {{\boldsymbol{h}}}^{\rm w}_{{\rm I}}} \right]_{r} \!
 {\beta_r}(t){e^{j{\varphi_r}(t)}}\!\right] = 0,\label{NLoSExp}
\end{alignat}
and their variances are given by
\begin{alignat}{1}
    \nonumber
    & {\rm{Var}}\!\left[\!{{\left[{ {\boldsymbol{g}}}^{{\rm{LOS}}}\right]_{r}\!
    \left[{ {{\boldsymbol{h}}}^{\rm w}_{{\rm I}}} \right]_{r} \!{\beta_r}(t){e^{j{\varphi_r}(t)}}} }\!\right] \\
    \nonumber
    & \;\;\;\;\;\;\;\;\;\;\;\;\;\;\;\;\;\;\;=  {\mathbb{E}}\!\left[ {{{\left|  \left[{ {{\boldsymbol{h}}}^{\rm w}_{{\rm I}}} \right]_{r}  \right|}^2}} \right] 
    {\mathbb{E}}\!\left[ {{{\left| {\beta_r}(t) \right|}^2}} \right],   \\ 
    & \;\;\;\;\;\;\;\;\;\;\;\;\;\;\;\;\;\;\;= \sum\limits_{i = 1}^{{2^b}} {{{\rm P}_i}\alpha _i^2} = \overline \alpha, \label{VarHACAele19}
\end{alignat}
and
\begin{alignat}{1}
    \nonumber
    & {\rm{Var}}\!\left[ \! \left[{ {\boldsymbol{g}}}^{{\rm{NLOS}}}\right]_{r}\!
    \left[{  {{\boldsymbol{h}}}^{\rm w}_{{\rm I}}} \right]_{r} \!
    {\beta_r}(t){e^{j{\varphi_r}(t)}}\!\right] \\
    \nonumber
    & \;\;\;\;\;\;\;\;\;\; =  {\mathbb{E}}\!\left[ {{{\left|   \left[{ {\boldsymbol{g}}}^{{\rm{NLOS}}}\right]_{r}  \right|}^2}} \right] 
    {\mathbb{E}}\!\left[ {{{\left|  \left[{ {{\boldsymbol{h}}}^{\rm w}_{{\rm I}}} \right]_{r}  \right|}^2}} \right] 
    {\mathbb{E}}\!\left[ {{{\left| {\beta_r}(t) \right|}^2}} \right],    \\ 
    & \;\;\;\;\;\;\;\;\;\; = \sum\limits_{i = 1}^{{2^b}} {{{\rm P}_i}\alpha _i^2} = \overline \alpha, \label{VarHACAele191}
\end{alignat}
where  ${{\rm P}_i}$ is the probability that the phase shift ${\varphi _r}(t)$ 
takes the $i$-th value of $\Phi$, i.e., ${{\rm P}_i} = \mathbb{P}\!\left({\varphi _r}(t)= {\phi _i}\right), \forall r$.

Since the number of DRIS  elements is large, 
according to the Lindeberg-L$\acute{e}$vy central limit theorem, 
$\frac{{h}^{\rm{w}}_{{\rm{D}}} (t)}{{{\mathcal{L}^{{{\frac{\nu_{{\rm{g}}}}{2}}}}}}{{\mathcal{L}^{{{\frac{\nu^{\rm{w}}_{{\rm{d}}}}{2}}}}}}}$ converges in distribution 
as follows:  
\begin{equation}
    \frac{{h}^{\rm{w}}_{{\rm{D}}} (t)}{{{\mathcal{L}^{{{\frac{\nu_{{\rm{g}}}}{2}}}}}}{{\mathcal{L}^{{{\frac{\nu^{\rm{w}}_{{\rm{d}}}}{2}}}}}}}
        \mathop  \to \limits^{\rm{d}} \mathcal{CN}\!\left( {0,  {  \frac{{N\!_{\rm D}}{\overline \alpha}}{{{\mathcal{L}^{{{{\nu_{{\rm{g}}}}}}}}}{{\mathcal{L}^{{{{\nu^{\rm{w}}_{{\rm{d}}}}}}}}}} } } \right).
        \label{HDStare18}
\end{equation}

\section{Proof of Proposition~\ref{Proposition11}}\label{AppendixA1}
According to~\eqref{ywPDFGua} and~\eqref{ywPDFH0}, 
we can obtain the following LRT:
\begin{alignat}{1}
\nonumber
\rho  & = \frac{\delta _{\rm{w}}^2}{\delta _{\rm{w}}^2 \!+\! P_0{{\left| \!{{\frac{{h^{\rm{w}}_{{\rm{d}}} } }{{{\mathcal{L}^{\frac{{{\nu^{\rm{w}}_{{\rm{d}}}}}}{2}}}}}}}
+ \! {{\frac{{h^{\rm{w}}_{{\rm{D}}} (m)} }{{  {\mathcal{L}^{\frac{{{\nu _{{\rm{g}}}}}}{2}}} {\mathcal{L}^{\frac{{{\nu^{\rm{w}}_{{\rm{I}}}}}}{2}}} } }}} \!\right|}^2}} \times \\
&\;\;\;\;\;\;\; \exp\!\!\left({\!\!{ - \frac{{\left| {{y_{\rm{w}}}\left( m \right)} \right|^2}}{\delta _{\rm{w}}^2 \!+\! P_0{{\left| \!{{\frac{{h^{\rm{w}}_{{\rm{d}}} } }{{{\mathcal{L}^{\frac{{{\nu^{\rm{w}}_{{\rm{d}}}}}}{2}}}}}}}
+ \! {{\frac{{h^{\rm{w}}_{{\rm{D}}} (m)} }{{  {\mathcal{L}^{\frac{{{\nu _{{\rm{g}}}}}}{2}}} {\mathcal{L}^{\frac{{{\nu^{\rm{w}}_{{\rm{I}}}}}}{2}}} } }}} \!\right|}^2}}} 
+ \frac{{\left| {{y_{\rm{w}}}\left( m \right)} \right|^2}}{\delta _{\rm{w}}^2}}\right)
\label{LRT10}
\end{alignat}
Setting $\rho = 1$, we then have 
\begin{alignat}{1}
\nonumber
{\left|y_{\rm w}(m) \right|^2} &=\frac{{ \left(\!{\delta _{\rm{w}}^2 \!+\! P_0{{\left| \!{{\frac{{h^{\rm{w}}_{{\rm{d}}} } }{{{\mathcal{L}^{\frac{{{\nu^{\rm{w}}_{{\rm{d}}}}}}{2}}}}}}}
+ \! {{\frac{{h^{\rm{w}}_{{\rm{D}}} (m)} }{{  {\mathcal{L}^{\frac{{{\nu _{{\rm{g}}}}}}{2}}} {\mathcal{L}^{\frac{{{\nu^{\rm{w}}_{{\rm{I}}}}}}{2}}} } }}} \!\right|}^2}} \right)\!
\delta _{\rm{w}}^2}}{{P_0{\left| \!{{\frac{{h^{\rm{w}}_{{\rm{d}}} } }{{{\mathcal{L}^{\frac{{{\nu^{\rm{w}}_{{\rm{d}}}}}}{2}}}}}}}
+\! {{\frac{{h^{\rm{w}}_{{\rm{D}}} (m)} }{{  {\mathcal{L}^{\frac{{{\nu _{{\rm{g}}}}}}{2}}} {\mathcal{L}^{\frac{{{\nu^{\rm{w}}_{{\rm{I}}}}}}{2}}} } }}} \!\right|}^2}} \times\\
&   \left(\!{\ln {\!\!\left(\!{\delta _{\rm{w}}^2 \!+\! P_0{{\left| \!{{\frac{{h^{\rm{w}}_{{\rm{d}}} } }{{{\mathcal{L}^{\frac{{{\nu^{\rm{w}}_{{\rm{d}}}}}}{2}}}}}}}
+ \! {{\frac{{h^{\rm{w}}_{{\rm{D}}} (m)} }{{  {\mathcal{L}^{\frac{{{\nu _{{\rm{g}}}}}}{2}}} {\mathcal{L}^{\frac{{{\nu^{\rm{w}}_{{\rm{I}}}}}}{2}}} } }}} \!\right|}^2}}\right)} - \ln {\delta^2_{\rm{w}}}}\!\right).
\label{OptimalywLRT}
\end{alignat}
Consequently, the optimal detection thresholds $\varepsilon(m) $ for the detection rule in~\eqref{decisionrule1} can be 
obtained as in~\eqref{OptimalLRT}.

\section{Proof of Theorem~\ref{Theorem11}}\label{AppendixB}
We first derive the FAR in~\eqref{FARExpre1}.
 When Alice and Bob are silent (${{\mathcal{H}}_0}$), the observations at Willie consist entirely of AWGN, 
i.e., ${\boldsymbol{y}_{\rm w}} = \left[y_{\rm w}(1),\cdots,y_{\rm w}(M)\right]^T = 
\left[n_{\rm w}(1),\cdots,n_{\rm w}(M)\right]^T$.
Therefore, the modulus squared of the $m$-th sample $\left|y_{\rm w}(m) \right|^2$,
follows the exponential distribution with PDF
\begin{equation}
    {f_{{{\left| {{Y_{\rm{w}} }} \right|}^2}}}\!\left( y \right) = \frac{1}{\pi{\delta _{\rm{w}}^2}}{e^{\frac{{ - y}}{{\delta _{\rm{w}}^2}}}}.
    \label{ProPym}
\end{equation}
Consequently, 
the cumulative distribution function (CDF) of $\left|y_{\rm w}(m) \right|^2$ under ${{\mathcal{H}}_0}$ can be expressed as
\begin{equation}
    {F_{{{\left| {{Y_{\rm{w}} }} \right|}^2}}}\!\left( y \right) 
    = 1- {e^{\frac{{ - y}}{{\delta _{\rm{w}}^2}}}}.
    \label{CDFym}
\end{equation}

 Under ${{\mathcal{H}}_0}$, the probability that 
there are only $T$ samples with power greater than the detection thresholds in~\eqref{OptimalLRT}
is calculated as
\begin{alignat}{1}
    \nonumber
     & {\left. {{p_T}} \right|{{\mathcal{H}}_0}} = {\mathbb{P}}\! \left( \left|{\boldsymbol{y}}_{\rm w}(i_1)\right|^2 \ge \varepsilon(i_1),\cdots, 
     \left|{\boldsymbol{y}}_{\rm w}(i_T)\right|^2 \ge \varepsilon(i_T) |{{\mathcal{H}}_0} \right) \\
    \nonumber
     & = \!\! {{\sum\limits_{\!{i_1} \!<  \cdots  < {i_T}} \!\!\! \left( \mathop \prod \limits_{j = {i_1}}^{{i_T}} \! {\mathbb{P}}\! \left( \!\left|{\boldsymbol{y}}_{\rm w}(j)\right|^2 \!\ge\! \varepsilon(j) |{{\mathcal{H}}_0} \! \right)
    \!  \right. }} \\
    & \;\;\;\;\;\;\;\;\;\;\;\;\;\;\;\;\;\;\;\;\;\; \left.\mathop \prod \limits_{i \ne {i_1} \ne  \cdots  \ne {i_T}}\!\!{\mathbb{P}}\! \left( \!\left|{\boldsymbol{y}}_{\rm w}(i)\right|^2 \!< \! \varepsilon(i) |{{\mathcal{H}}_0} \! \right)\!\!\right),
    \label{COmT1}
\end{alignat}
where $1 \le i_1 < \cdots < i_T \le M$.

According to~\eqref{ProPym} and~\eqref{CDFym}, the probability ${\left. {{p_T}} \right|{{\mathcal{H}}_0}}$ can be represented as
\begin{equation}
    {\left. {{p_T}} \right|{{\mathcal{H}}_0}} =  {{\sum\limits_{{i_1} <  \cdots  < {i_T}} \!\! 
    \!\left({{\mathop \prod \limits_{j = {i_1}}^{{i_T}} \!e^{\frac{{ - \varepsilon(j)}}{{\delta _{\rm{w}}^2}}}} \mathop \prod \limits_{i \ne {i_1} \ne  \cdots  \ne {i_T}}\!\!\!\! 
    \left(\! {1 - e^{- \frac{{\varepsilon (i)}}{{\delta _{\rm{w}}^2}}}} \!\right)}\!\!\right) }}. \label{COmT2} 
\end{equation}
The FAR of Willies detection test can be computed as
\begin{equation}
    p_{\rm{F}} =   \sum\limits_{T = N}^M {{\left. {{p_T}} \right|{{\mathcal{H}}_0}}}.
    \label{FARexpxx}
\end{equation}
Substituting~\eqref{COmT2} to~\eqref{FARexpxx}, the FAR in~\eqref{FARExpre1} is obtained.

On the other hand, the MDR in~\eqref{MDRExpre} corresponds to the case where Alice and Bob are transmitting (${{\mathcal{H}}_1}$),
but the warden Willie fails to detect the transmission.
Based on~\eqref{ywPDFGua}, $\left|y_{\rm w}(m) \right|^2$ follows the exponential PDF given by
\begin{alignat}{1}
    \nonumber
    & {f_{{{\left| {{Y_{\rm{w}} }} \right|}^2}}}\!\left( y \right) = \\
   &  \frac{1}{\pi\! \!\left(\! {{\delta _{\rm{w}}^2 \!+\! P_0{{\left| \!{{\frac{{h^{\rm{w}}_{{\rm{d}}} } }{{{\mathcal{L}^{\frac{{{\nu^{\rm{w}}_{{\rm{d}}}}}}{2}}}}}}}
   \!+ \!\! {{\frac{{h^{\rm{w}}_{{\rm{D}}} (m)} }{{  {\mathcal{L}^{\frac{{{\nu _{{\rm{g}}}}}}{2}}} \!{\mathcal{L}^{\frac{{{\nu^{\rm{w}}_{{\rm{I}}}}}}{2}}} } }}} \!\right|}^2}}} \right)}
     {\exp\!\!\left(\!\!{\frac{{ - y}}{{\delta _{\rm{w}}^2 \!+\! P_0{{\left| \!{{\frac{{h^{\rm{w}}_{{\rm{d}}} } }{{{\mathcal{L}^{\frac{{{\nu^{\rm{w}}_{{\rm{d}}}}}}{2}}}}}}}
     \!+\! \! {{\frac{{h^{\rm{w}}_{{\rm{D}}} (m)} }{{  {\mathcal{L}^{\frac{{{\nu _{{\rm{g}}}}}}{2}}} \!{\mathcal{L}^{\frac{{{\nu^{\rm{w}}_{{\rm{I}}}}}}{2}}} } }}} \!\right|}^2}}}}\!\!\right)}.
    \label{ProPym2nd}
\end{alignat}
According to~\eqref{ProPym2nd}, 
the CDF of $\left|y_{\rm w}(m) \right|^2$ under ${{\mathcal{H}}_1}$ can expressed as
\begin{equation}
    {F_{{{\left| {{Y_{\rm{w}} }} \right|}^2}}}\!\left( y \right) 
    = 1- {\exp\!\!\left(\!{\frac{{ - y}}{{{\delta _{\rm{w}}^2 \!+\! P_0{{\left| \!{{\frac{{h^{\rm{w}}_{{\rm{d}}} } }{{{\mathcal{L}^{\frac{{{\nu^{\rm{w}}_{{\rm{d}}}}}}{2}}}}}}}
    + \! {{\frac{{h^{\rm{w}}_{{\rm{D}}} (m)} }{{  {\mathcal{L}^{\frac{{{\nu _{{\rm{g}}}}}}{2}}} {\mathcal{L}^{\frac{{{\nu^{\rm{w}}_{{\rm{I}}}}}}{2}}} } }}} \!\right|}^2}}}}}\!\right)}.
    \label{CDFymH1}
\end{equation}

Under ${{\mathcal{H}}_1}$, the probability that 
there are only $T$ samples with power greater than the detection thresholds in~\eqref{OptimalLRT}
is given by
\begin{equation}
      {\left. {{p_T}} \right|{{\mathcal{H}}_1}} =  
      {{\sum\limits_{{i_1} <  \cdots  < {i_T}}   
     {\mathop \prod \limits_{j = {i_1}}^{{i_T}} \!{\exp\!\!\left(\!{\frac{{ - \varepsilon(j)}}{{{{\delta _{\rm{w}}^2 + \frac{{{P_0}{{\left| {h_{\rm{d}}^{\rm{w}}} \right|}^2}}}{{{\mathcal{L}^{\nu_{\rm{d}}^{\rm{w}}}}}} + 
     \frac{{{P_0}{{\left| {h_{\rm{D}}^{\rm{w}}(j)} \right|}^2}}}{{{\mathcal{L}^{{\nu_{\rm{g}}}}}{\mathcal{L}^{\nu_{\rm{I}}^{\rm{w}}}}}}}}}}}\!\right)}    } }} . \label{COmT22} 
\end{equation}
Consequently, the MDR for Willie's detection is calculated as
\begin{equation}
    p_{\rm{M}} = {{\left. {{p_0}} \right|{{\mathcal{H}}_1}}} + \sum\limits_{T = 1}^{N-1} {{\left. {{p_T}} \right|{{\mathcal{H}}_1}}},
    \label{MDRexpxx}
\end{equation}
where
\begin{equation}
    {\left. {{p_0}} \right|{{\mathcal{H}}_1}} = { \mathop \Pi \limits_{m = 1}^M 
    {\!\!\left(\!\!1-\exp\!\left(\!{-\frac{\varepsilon(m)}{{{\delta _{\rm{w}}^2 + \frac{{{P_0}{{\left| {h_{\rm{d}}^{\rm{w}}} \right|}^2}}}{{{\mathcal{L}^{\nu_{\rm{d}}^{\rm{w}}}}}} + 
    \frac{{{P_0}{{\left| {h_{\rm{D}}^{\rm{w}}(m)} \right|}^2}}}{{{\mathcal{L}^{{\nu_{\rm{g}}}}}{\mathcal{L}^{\nu_{\rm{I}}^{\rm{w}}}}}}}}}}\!\!\right)\!\!\right)} } .
    \label{MDRexp0}
\end{equation}
As a result, the MDR is derived, as given in~\eqref{MDRExpre}.

\section{Proof of Theorem~\ref{Theorem2}}\label{AppendixC}
Similar to the conclusion of Proposition~\ref{Proposition1},
the cascaded DRIS-jammed channel $h_{\rm D}^{b}(t)$ also converges
in distribution to a complex Gaussian random variable as the number of DRIS elements 
grows large:
\begin{equation}
    \frac{{h}^{\rm{b}}_{{\rm{D}}}(t) }{{{\mathcal{L}^{{{\frac{\nu_{{\rm{g}}}}{2}}}}}}{{\mathcal{L}^{{{\frac{\nu^{\rm{b}}_{{\rm{d}}}}{2}}}}}}}
    \mathop  \to \limits^{\rm{d}} \mathcal{CN}\!\!\left(\! {0,  {  \frac{{N\!_{\rm D}}{\overline \alpha}}{{{\mathcal{L}^{{{{\nu_{{\rm{g}}}}}}}}}{{\mathcal{L}^{{{{\nu^{\rm{b}}_{{\rm{d}}}}}}}}}} } } \right), 
    N_{\rm D} \to \infty.
    \label{HdSta}
\end{equation}
Conditioned on the fact the covert message $s(m)$ and the DRIS-jammed channel are independent, 
the expectation in~\eqref{eqSLNR} can be reduced to
\begin{equation}
    \mathbb{E} \!\left[\left|{h^{\rm{b}}_{{\rm{D}}}(m)}s(m)\right|^2\right] = 
    \mathbb{E} \!\left[\left|{h^{\rm{b}}_{{\rm{D}}}(m)} \right|^2\right] \mathbb{E} \!\left[\left| s(m)\right|^2\right].
    \label{HDdsm}
\end{equation}
Based on~\eqref{HdSta}, the DRIS-based ACA interference in~\eqref{eqSLNR} converges in distribution to a fixed value 
\begin{equation}
    { \mathbb{E} \!\left[\left|{h^{\rm{b}}_{{\rm{D}}}(m)}s(m)\right|^2\right] } 
    \mathop  \to \limits^{\rm{d}} P_0 N_{\rm D}{\overline \alpha}, 
    \;{\rm{as}}\; N_{\rm D} \to \infty.
    \label{HdACASta}
\end{equation}  

Since ${h^{\rm{b}}_{{\rm{d}}}}$ and $s(m)$ are independent random variables, the expectation of $\left|{h^{\rm{b}}_{{\rm{d}}}}s(m)\right|^2$ can  be 
simplified to
\begin{equation}
    { \mathbb{E} \!\left[\left|{h^{\rm{b}}_{{\rm{d}}}}s(m)\right|^2\right] } 
    ={ \mathbb{E} \!\left[\left|{h^{\rm{b}}_{{\rm{d}}}}\right|^2\right] } { \mathbb{E} \!\left[\left|s(m)\right|^2\right] } .
    \label{HdbsSta}
\end{equation} 
Based on the definition in~\eqref{Hdbkeq}, we have 
\begin{equation}
    { \mathbb{E} \!\left[\left|{h^{\rm{b}}_{{\rm{d}}}s(m)}\right|^2\right] } 
    =P_0.
    \label{HdbSta}
\end{equation} 

Substituting~\eqref{HdACASta} and~\eqref{HdbSta} to~\eqref{eqSLNR},
the ergodic SJNR can be obtained as given in~\eqref{SNRBobVau}.
\end{appendices}

\end{document}